%% file: root.tex
\documentclass[conference]{IEEEtran}
\IEEEoverridecommandlockouts
\usepackage[OT1]{fontenc} 
\usepackage{cite}
\usepackage{amsmath,amssymb,amsfonts}
\usepackage{textcomp}
\usepackage{xcolor}
\usepackage{blindtext}
\usepackage{hyperref}
\usepackage{subfiles} 
\usepackage{mathtools}
\usepackage{physics}
\usepackage{tikz}
\usepackage{mathdots}
\usepackage{yhmath}
\usepackage{cancel}
\usepackage{color}
\usepackage{siunitx}
\usepackage{array}
\usepackage{multirow}
\usepackage{amssymb}
\usepackage{gensymb}
\usepackage{adjustbox}
\usepackage{tabularx}
\usepackage{extarrows}
\usepackage{booktabs}
\usepackage{diagbox}
\usepackage{graphicx}
\usepackage[ruled,vlined]{algorithm2e}
\usepackage[normalem]{ulem}
\usepackage{subcaption}
\useunder{\uline}{\ul}{}
\usetikzlibrary{fadings}
\usetikzlibrary{patterns}
\usetikzlibrary{shadows.blur}
\usetikzlibrary{shapes}
\graphicspath{ {./Figures/} }

\DeclareMathOperator*{\argmin}{argmin}

\def\BibTeX{{\rm B\kern-.05em{\sc i\kern-.025em b}\kern-.08em
    T\kern-.1667em\lower.7ex\hbox{E}\kern-.125emX}}

\begin{document}

\title{\LARGE \bf Sequential Cooperative Energy and Time-Optimal Lane Change Maneuvers for Highway Traffic 
\thanks{Supported by the Honda Research Institute USA (HRI-USA), by NSF under grants ECCS-1931600,
DMS-1664644, CNS-1645681, by AFOSR under grant FA9550-19-1-0158,
by ARPA-E under grant DE-AR0001282, and by the MathWorks}
}

\author{Andres S. Chavez Armijos$^{1}$, Rui Chen$^{1}$, Christos G. Cassandras$^{1}$,
\thanks{$^{1}$A. S. Chavez Armijos, R. Chen, and C. G. Cassandras are
with the Division of Systems Engineering and Center for Information and
Systems Engineering, Boston University, Brookline, MA 02446
(email:\{aschavez; ruic; cgc\}@bu.edu).}
Yasir K. Al-Nadawi$^{2}$, \\Hossein Nourkhiz Mahjoub$^{2}$, and Hidekazu Araki$^{2}$
\thanks{$^{2}$Y. K. Al-Nadawi, H. Nourkhiz Mahjoub, and H. Araki are with Honda Research Institute-US (HRI-US)
Ann Arbor, MI 48103 USA
(email:\{yasir\_alnadawi; hnourkhizmahjoub; haraki\}@honda-ri.com)}
}

\maketitle
\begin{abstract}
\subfile{sections/abstract}
\end{abstract}

\section{Introduction}
\subfile{sections/introduction}
\section{Problem Formulation}\label{secII:ProblemFormulation}
\subfile{sections/problem_formulation}

\section{Decentralized Optimal Control Solution}\label{secIII:DecentralizedOCSol}
\subfile{sections/optimal_control_solution}

\section{Simulation Results}
\subfile{sections/simulation_results}

\section{Conclusions and Future Work}
\subfile{sections/Conclusions}

\bibliographystyle{IEEEtran}
\begin{tiny}
\bibliography{bibliography,cmp}
\end{tiny}

\end{document}

%% file: sections/abstract.tex
We derive optimal control policies for a Connected Automated Vehicle (CAV) and cooperating neighboring CAVs to carry out a lane change maneuver consisting of a longitudinal phase where the CAV properly positions itself relative to the cooperating neighbors and a lateral phase where
it safely changes lanes. 
In contrast to prior work on this problem, where the CAV ``selfishly'' seeks to minimize its maneuver time, we seek to ensure that the fast-lane traffic flow is minimally disrupted (through a properly defined metric) and that highway throughput is improved by optimally selecting the cooperating vehicles. We show that analytical solutions for the optimal trajectories can be derived and are guaranteed to satisfy safety constraints for all vehicles involved in the maneuver. When feasible solutions do not exist, we include a time relaxation method trading off a longer maneuver time with reduced disruption. Our analysis is also extended to multiple sequential maneuvers. Simulation results where the controllers are implemented show their effectiveness in terms of safety guarantees and up to 35\% throughput improvement compared to maneuvers with no vehicle cooperation.

%% file: sections/introduction.tex
Advances in transportation system technologies and the emergence of Connected
Automated Vehicles (CAVs), also known as \textquotedblleft autonomous
vehicles\textquotedblright, have the potential to drastically improve a
transportation network's performance in terms of safety, comfort, congestion
reduction and energy efficiency. In highway driving, an overview of automated
intelligent vehicle-highway systems was provided in \cite{varaiya1993smart}
with more recent developments mostly focusing on autonomous car-following
control given in \cite{zhao2018accelerated},\cite{wang2016cooperative}%
,\cite{wang2015game}. In urban driving, efforts have concentrated on
controlling traffic lights \cite{fleck2015adaptive} or the cooperative control
of CAVs crossing non-signalized intersections \cite{dresner2008multiagent}%
,\cite{zhang2019decentralized}.

Automating a lane change maneuver remains a challenging problem that has
attracted increasing attention in recent years \cite{nilsson2015longitudinal}%
,\cite{bax2014road},\cite{you2015trajectory},\cite{werling2010optimal}.
Designing such an automated maneuver is often viewed as consisting of two
levels \cite{bevly2016lane}: at the strategy level, a feasible trajectory is generated for lane changing; then, the
control level determines how vehicles track the
aforementioned trajectory.  The work in \cite{you2015trajectory} adopts such a design architecture for an automated lane-change maneuver, but does not
provide an analytical solution and assumes that there are no other vehicles in
the left lane (the lane in which the controllable vehicle ends up after
completing the maneuver). In \cite{nilsson2017lane}, background vehicles are
included in the left lane and the goal is to check whether there exists a
feasible lane-change trajectory or not; if one exists, the controllable vehicle will then track this trajectory. In these papers, only one vehicle
can be controlled during the maneuver and no analytical solutions are provided.

The emergence of CAVs creates the opportunity for cooperation among vehicles
traveling in multi-lane roads to carry out automated
lane-change maneuvers \cite{mahjoub2017learning}
\cite{desiraju2014minimizing}, \cite{luo2016dynamic}, \cite{li2020cooperative}. Such cooperation offers several advantages
relative to the two-level architecture mentioned above. In particular, when
controlling a single vehicle and checking on the feasibility of a maneuver
depending on the state of nearby traffic, as in \cite{kamal2013model}%
,\cite{katriniok2013optimal}, the maneuver is often infeasible without the
cooperation of other vehicles, especially under heavier traffic conditions. In
contrast, a cooperative approach can allow multiple interacting vehicles to
implement controllers enabling a larger set of feasible maneuvers. Aside from enhancing
safety, this cooperative behavior can also improve the throughput, if designed properly, hence reducing the chance of congestion. 

The problem of cooperative multi-agent lane-changing maneuvers can be solved as a centralized or decentralized motion planning problem. In the centralized case, a Control Zone (CZ) is defined as a prespecified area within which all vehicles follow commands issued by a central coordinator (e.g. traffic beacon) that computes a solution for every vehicle involved. In contrast, in the decentralized case, the multi-agent problem is solved by each individual agent computing their individual solution on-board. In the centralized approach, the computation can become intractable and requires robust communication guarantees. In the decentralized case, on the other hand, solutions may be too conservative and could generate unwanted disruptions in traffic  \cite{li2018balancing}.

Feasible, but not necessarily optimal,
vehicle trajectories for cooperative multi-agent lane-changing maneuvers are
derived in \cite{lam2013cooperative}. The case of multiple cooperating
vehicles simultaneously changing lanes is considered in \cite{li2017optimal}
with the requirement that all vehicles are controllable and their velocities
before the lane change are all the same. First, vehicles with a lower
priority must adjust their positions in their current lane and give way to
those with a higher priority to avoid collisions. Then, an optimal control problem (OCP) is solved for each vehicle without considering the
usual safe distance constraints between vehicles. This \textquotedblleft
progressively constrained dynamic optimization\textquotedblright\ method
facilitates a numerical solution to the underlying OCP at
the expense of some loss in performance.

\begin{figure} [pt]
    \centering
    \vspace*{\baselineskip}
    \begin{tikzpicture}[x=0.75pt,y=0.75pt,yscale=-0.66,xscale=0.44]
        \draw [line width=3]    (0,35.16) -- (358.76,35.7) -- (740,35.7) ;
        \draw [line width=3]    (0,239.03) -- (321.62,239.03) -- (740,239.03) ;
        \draw [color={rgb, 255:red, 248; green, 231; blue, 28 }  ,draw opacity=1 ][fill={rgb, 255:red, 248; green, 231; blue, 28 }  ,fill opacity=1 ][line width=3]  [dash pattern={on 13pt off 10pt}]  (0,138.11) -- (740.81,136.69) ;
        \draw (500,185) node  {\includegraphics[scale=0.11]{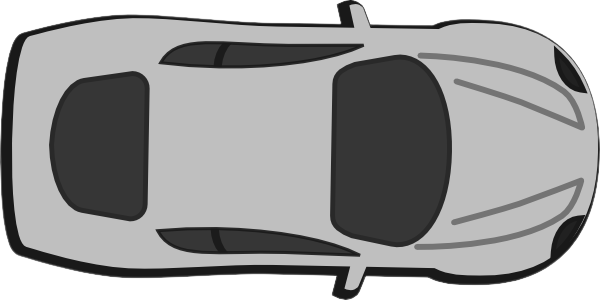}};
        \draw (212.09,185) node  {\includegraphics[scale=0.07]{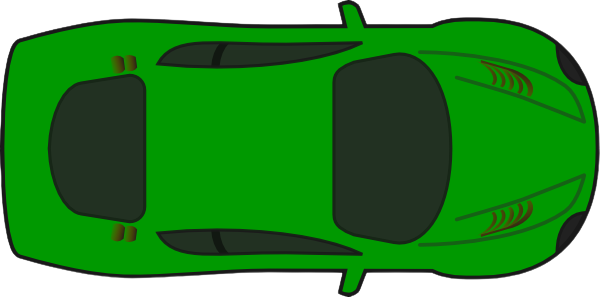}};
        \draw (514.97,90) node  {\includegraphics[scale=0.17]{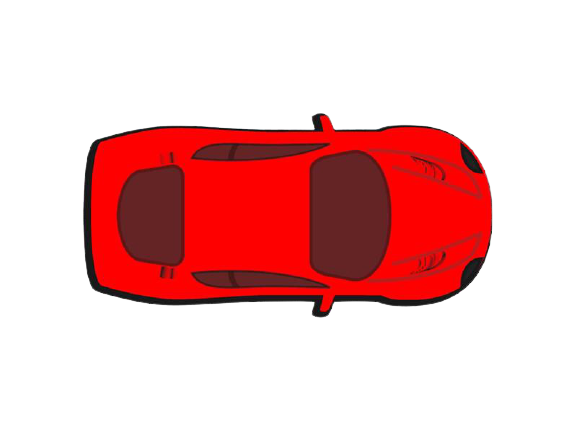}};
        \draw (195.97,90) node  {\includegraphics[scale=0.17]{redVehicleTopView.png}};
        \draw [color={rgb, 255:red, 65; green, 117; blue, 5 }  ,draw opacity=1 ] [dash pattern={on 3.75pt off 3pt on 7.5pt off 1.5pt}]  (270.18,184.81) .. controls (368.48,192.9) and (296.46,68) .. (417.56,80.51) ;
        \draw [shift={(419.4,80.71)}, rotate = 186.44] [fill={rgb, 255:red, 65; green, 117; blue, 5 }  ,fill opacity=1 ][line width=0.08]  [draw opacity=0] (10.72,-5.15) -- (0,0) -- (10.72,5.15) -- (7.12,0) -- cycle    ;
        \draw [color={rgb, 255:red, 208; green, 2; blue, 27 }  ,draw opacity=1 ][fill={rgb, 255:red, 208; green, 2; blue, 27 }  ,fill opacity=1 ] [dash pattern={on 3.75pt off 3pt on 7.5pt off 1.5pt}]  (114.82,90) -- (81.38,90) ;
        \draw [shift={(79.38,90)}, rotate = 360] [color={rgb, 255:red, 208; green, 2; blue, 27 }  ,draw opacity=1 ][line width=0.75]    (10.93,-3.29) .. controls (6.95,-1.4) and (3.31,-0.3) .. (0,0) .. controls (3.31,0.3) and (6.95,1.4) .. (10.93,3.29)   ;
        \draw [color={rgb, 255:red, 208; green, 2; blue, 27 }  ,draw opacity=1 ] [dash pattern={on 3.75pt off 3pt on 7.5pt off 1.5pt}]  (576.12,90) -- (620.51,90) ;
        \draw [shift={(620.51,90)}, rotate = 180] [color={rgb, 255:red, 208; green, 2; blue, 27 }  ,draw opacity=1 ][line width=0.75]    (10.93,-3.29) .. controls (6.95,-1.4) and (3.31,-0.3) .. (0,0) .. controls (3.31,0.3) and (6.95,1.4) .. (10.93,3.29)   ;
        
        \draw (0,45) node [anchor=north west][inner sep=0.3pt]  [font=\large] [align=left] {$\displaystyle u_{2}^{*} \leq 0$$ $};
        \draw (592.81,48.15) node [anchor=north west][inner sep=0.75pt]  [font=\large] [align=left] {$\displaystyle u_{1}^{*} \geq 0$$ $};
        \draw (335.63,145.53) node [anchor=north west][inner sep=0.75pt]  [font=\large] [align=left] {$\displaystyle u_{C}^{*}$$ $};
        \draw (215.95,183) node  [font=\large] [align=left] {\begin{minipage}[lt]{26.64pt}\setlength\topsep{0pt}
        C
        \end{minipage}};
        \draw (510.26,183) node  [font=\large] [align=left] {\begin{minipage}[lt]{26.64pt}\setlength\topsep{0pt}
        U
        \end{minipage}};
        \draw (525.44,90) node  [font=\large] [align=left] {\begin{minipage}[lt]{26.64pt}\setlength\topsep{0pt}
        1
        \end{minipage}};
        \draw (210,90) node  [font=\large] [align=left] {\begin{minipage}[lt]{26.64pt}\setlength\topsep{0pt}
        2
        \end{minipage}};
    \end{tikzpicture}
    \caption{The basic lane-changing maneuver process.}
    \label{Fig1:original_maneuver_diagram}
    \vspace*{-\baselineskip}
\end{figure}

In our previous work, \cite{chen2020cooperative} we provided a time and energy optimal analytical solution for the maneuver shown in
Fig. \ref{Fig1:original_maneuver_diagram}, in which the controlled vehicle $C$ attempts to
overtake an uncontrollable vehicle $U$ by using the left lane to pass. 
A decentralized analytical solution is provided based on the cooperation and communication with two neighboring vehicles (vehicles $1$ and $2$) with the goal of minimizing the total maneuver time and subsequently determining trajectories that minimize the energy consumed by all three cooperating vehicles. 
This approach applies to a wider range of scenarios
relative to those in \cite{nilsson2017lane}, \cite{luo2016dynamic}%
, \cite{kamal2013model}, \cite{katriniok2013optimal} and incorporates the safety
distance constraint not included in \cite{lam2013cooperative} and
\cite{li2017optimal}. 
However, by seeking to minimize $C$'s maneuver time, \cite{chen2020cooperative} adopts a \emph{vehicle-centric} (selfish) viewpoint which ignores the effect of the maneuver on all remaining vehicles. As a result, since vehicle $2$ typically decelerates to allow $C$ to get ahead of it, this deceleration may cause a traffic flow slowdown in the left lane which can negatively impact throughput, especially when traffic is relatively heavy. Moreover, the analysis assumes that vehicles $1$ and $2$ are predetermined rather than being optimally selected among a set of possible cooperation candidates.

The contribution of this paper is to alleviate both of these limitations in such ``\emph{selfish}'' maneuvers by adopting instead a ``\emph{system-centric}'' optimality viewpoint. This provides a decentralized but socially optimal solution in the sense that
our optimal controller design ensures that the resulting traffic throughput on the highway is improved by adding vehicles to the fast lane while (i) limiting the ``disruption'' that cooperation among multiple vehicles on the road can cause on the fast lane traffic flow, and (ii) determining an optimal pair of cooperating vehicles, which play the role of $1$ and $2$ in Fig. \ref{Fig1:original_maneuver_diagram}, within a set of feasible such candidates.
Additionally, we focus on the realization of \emph{multiple} sequential lane-changing maneuvers under the assumption of cooperation allowance by surrounding vehicles. As in \cite{chen2020cooperative}, we decompose the maneuver into a longitudinal component followed by a lateral component. In the longitudinal part, our approach is based on first determining an optimal maneuver time for $C$ subject to all safety and speed and acceleration constraints for vehicles $C$,
$1$, and $2$ and such that $C$ attains a desired final speed that matches that of the left lane traffic flow (if possible). We then solve a fixed terminal time decentralized optimal control
problem for each of the two cooperating vehicles in which energy
consumption is minimized. In the lateral phase, we solve a decentralized optimal control
problem seeking to jointly minimize the time and energy consumed which is no different than the one presented in \cite{chen2020cooperative}.
Our analysis also allows the determination of a vehicle pair that results in minimal acceleration/deceleration for this pair. This minimizes the possibility of excessive braking or deceleration of the rear vehicle in the pair ($2$ in Fig. \ref{Fig1:original_maneuver_diagram}), quantified through an appropriate ``disruption metric''.
Our results show that vehicles form natural platoons that dictate the free-flow speed of the fast lane in a two-lane highway. 

The rest of the paper is organized as follows. Section II presents the formulation of the
longitudinal lane-change maneuver problem. In Section III,
a complete optimal control solution to coordinate multiple lane change maneuvers is obtained.
Section IV provides simulation results for
several representative examples and we conclude with Section VI.

%% file: sections/problem_formulation.tex
In this section, we formulate the system-centric version of the cooperative maneuver setting presented in \cite{chen2020cooperative} which was exclusively vehicle-centric. As in \cite{chen2020cooperative}, we decompose the maneuver into a longitudinal component followed by a lateral component. The former is significantly different, including the determination of a minimally disrupting cooperative pair (playing the role of $1$ and $2$ in Fig. \ref{Fig1:original_maneuver_diagram}). However, the latter is the same and we omit details that are given in \cite{chen2020cooperative}.

\begin{figure}[pt]
    \centering
    \vspace*{\baselineskip}
    \tikzset{every picture/.style={line width=0.75pt}} 
        \begin{tikzpicture}[x=0.75pt,y=0.75pt,yscale=-1,xscale=0.83]
        
        \draw [line width=3]    (50.81,39.52) -- (239.1,39.21) -- (445.37,39.35) ;
        \draw [line width=3]    (50.43,130.42) -- (218.61,131.02) -- (444.99,130.59) ;
        \draw [color={rgb, 255:red, 248; green, 231; blue, 28 }  ,draw opacity=1 ][fill={rgb, 255:red, 248; green, 231; blue, 28 }  ,fill opacity=1 ][line width=3]  [dash pattern={on 11.25pt off 9.75pt}]  (51.33,89.28) -- (446,88.77) ;
        \draw (343.84,104.67) node [rotate=-0.02,xslant=0] {\includegraphics[width=26.23pt,height=16.6pt]{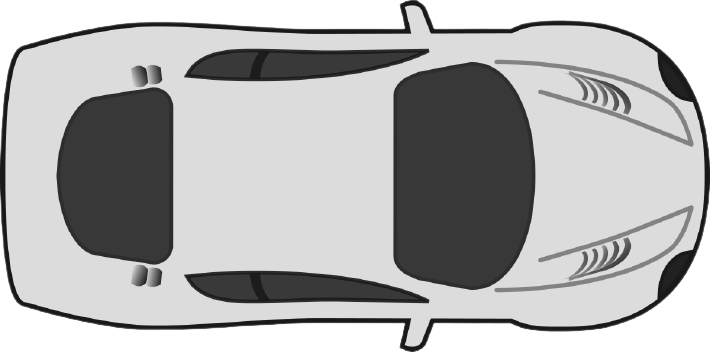}};
        \draw (284.48,61.06) node [rotate=-0.02,xslant=0] {\includegraphics[width=26.23pt,height=16.6pt]{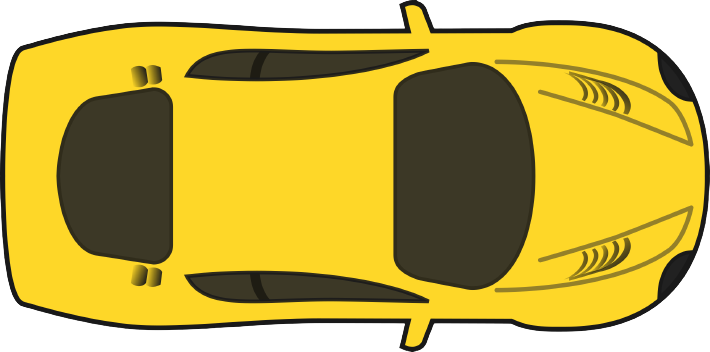}};
        \draw (421.59,61.13) node [rotate=-0.02,xslant=0] {\includegraphics[width=26.23pt,height=16.6pt]{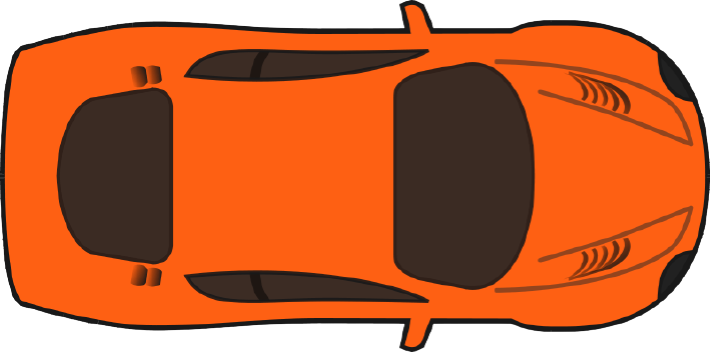}};
        \draw (361.04,61.1) node [rotate=-0.02,xslant=0] {\includegraphics[width=26.23pt,height=16.6pt]{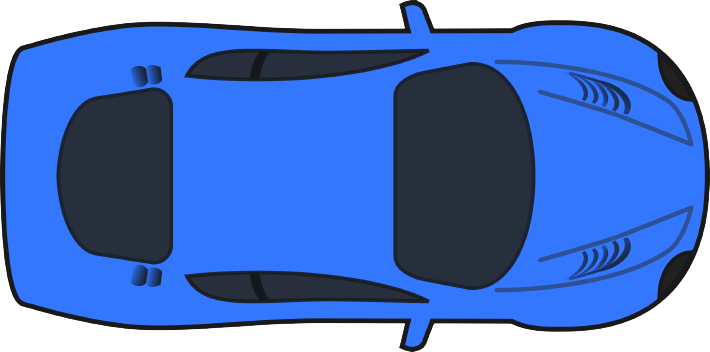}};
        \draw (260.03,104.17) node [rotate=-0.02,xslant=0] {\includegraphics[width=26.23pt,height=16.6pt]{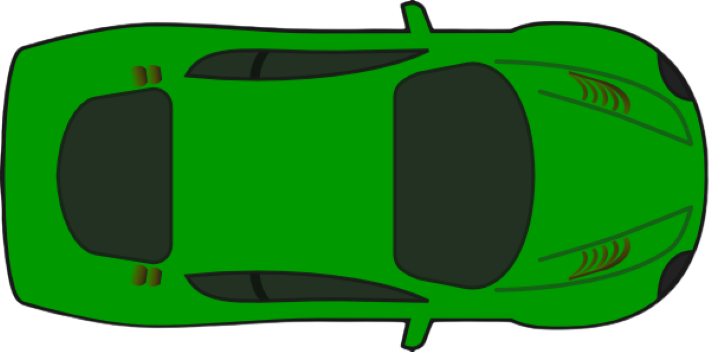}};
        \draw (214.3,60.57) node [rotate=-0.02,xslant=0] {\includegraphics[width=26.23pt,height=16.6pt]{car_yellow.png}};
        \draw (144.84,61.02) node [rotate=-0.02,xslant=0] {\includegraphics[width=26.23pt,height=16.6pt]{car_blue.png}};
        \draw (75.41,60.96) node [rotate=-0.02,xslant=0] {\includegraphics[width=26.23pt,height=16.6pt]{car_orange.png}};
        \draw (156.21,105.55) node [rotate=-0.02,xslant=0] {\includegraphics[width=26.23pt,height=16.6pt]{car_orange.png}};
        \draw  [color={rgb, 255:red, 208; green, 2; blue, 27 }  ,draw opacity=1 ][dash pattern={on 4.5pt off 4.5pt}] (111.4,51.5) .. controls (111.41,46.8) and (115.22,42.99) .. (119.92,42.99) -- (386.18,43.09) .. controls (390.88,43.09) and (394.69,46.9) .. (394.68,51.6) -- (394.66,77.11) .. controls (394.65,81.81) and (390.84,85.62) .. (386.14,85.62) -- (119.87,85.53) .. controls (115.18,85.52) and (111.37,81.71) .. (111.38,77.02) -- cycle ;
        \draw [color={rgb, 255:red, 155; green, 155; blue, 155 }  ,draw opacity=1 ][fill={rgb, 255:red, 155; green, 155; blue, 155 }  ,fill opacity=1 ] [dash pattern={on 3.75pt off 3pt on 7.5pt off 1.5pt}]  (239.09,103.67) -- (239,145.95) ;
        \draw [color={rgb, 255:red, 155; green, 155; blue, 155 }  ,draw opacity=1 ][fill={rgb, 255:red, 155; green, 155; blue, 155 }  ,fill opacity=1 ] [dash pattern={on 3.75pt off 3pt on 7.5pt off 1.5pt}]  (175.18,145.92) -- (113.32,145.89) ;
        \draw [shift={(111.32,145.89)}, rotate = 0.02] [color={rgb, 255:red, 155; green, 155; blue, 155 }  ,draw opacity=1 ][line width=0.75]    (10.93,-3.29) .. controls (6.95,-1.4) and (3.31,-0.3) .. (0,0) .. controls (3.31,0.3) and (6.95,1.4) .. (10.93,3.29)   ;
        \draw [color={rgb, 255:red, 155; green, 155; blue, 155 }  ,draw opacity=1 ][fill={rgb, 255:red, 155; green, 155; blue, 155 }  ,fill opacity=1 ] [dash pattern={on 3.75pt off 3pt on 7.5pt off 1.5pt}]  (175.25,145.92) -- (237.47,145.95) ;
        \draw [shift={(239.47,145.95)}, rotate = 180.02] [color={rgb, 255:red, 155; green, 155; blue, 155 }  ,draw opacity=1 ][line width=0.75]    (10.93,-3.29) .. controls (6.95,-1.4) and (3.31,-0.3) .. (0,0) .. controls (3.31,0.3) and (6.95,1.4) .. (10.93,3.29)   ;
        \draw [color={rgb, 255:red, 155; green, 155; blue, 155 }  ,draw opacity=1 ][fill={rgb, 255:red, 155; green, 155; blue, 155 }  ,fill opacity=1 ] [dash pattern={on 3.75pt off 3pt on 7.5pt off 1.5pt}]  (111.39,62.19) -- (111.32,145.89) ;
        
        \draw [color={rgb, 255:red, 155; green, 155; blue, 155 }  ,draw opacity=1 ] [dash pattern={on 3.75pt off 3pt on 7.5pt off 1.5pt}]  (368.31,104.71) -- (368.29,146.25) ;
        \draw [color={rgb, 255:red, 155; green, 155; blue, 155 }  ,draw opacity=1 ] [dash pattern={on 3.75pt off 3pt on 7.5pt off 1.5pt}]  (381.44,146.26) -- (392.6,146.26) ;
        \draw [shift={(394.6,146.26)}, rotate = 180.02] [color={rgb, 255:red, 155; green, 155; blue, 155 }  ,draw opacity=1 ][line width=0.75]    (10.93,-3.29) .. controls (6.95,-1.4) and (3.31,-0.3) .. (0,0) .. controls (3.31,0.3) and (6.95,1.4) .. (10.93,3.29)   ;
        \draw [color={rgb, 255:red, 155; green, 155; blue, 155 }  ,draw opacity=1 ] [dash pattern={on 3.75pt off 3pt on 7.5pt off 1.5pt}]  (381.42,146.26) -- (370.19,146.25) ;
        \draw [shift={(368.19,146.25)}, rotate = 0.02] [color={rgb, 255:red, 155; green, 155; blue, 155 }  ,draw opacity=1 ][line width=0.75]    (10.93,-3.29) .. controls (6.95,-1.4) and (3.31,-0.3) .. (0,0) .. controls (3.31,0.3) and (6.95,1.4) .. (10.93,3.29)   ;
        \draw [color={rgb, 255:red, 155; green, 155; blue, 155 }  ,draw opacity=1 ] [dash pattern={on 3.75pt off 3pt on 7.5pt off 1.5pt}]  (394.67,64.01) -- (394.6,146.26) ;
        
        \draw  [dash pattern={on 0.75pt off 0.75pt}]  (278.37,103.85) .. controls (280.06,102.2) and (281.72,102.22) .. (283.37,103.91) -- (287.48,103.97) -- (295.48,104.07) ;
        \draw [shift={(297.48,104.09)}, rotate = 180.73] [color={rgb, 255:red, 0; green, 0; blue, 0 }  ][line width=0.75]    (10.93,-3.29) .. controls (6.95,-1.4) and (3.31,-0.3) .. (0,0) .. controls (3.31,0.3) and (6.95,1.4) .. (10.93,3.29)   ;
        \draw  [color={rgb, 255:red, 245; green, 166; blue, 35 }  ,draw opacity=1 ][dash pattern={on 4.5pt off 4.5pt}] (184.09,52.4) .. controls (184.1,48.16) and (187.53,44.73) .. (191.77,44.73) -- (303.01,44.77) .. controls (307.24,44.77) and (310.68,48.21) .. (310.67,52.44) -- (310.65,75.45) .. controls (310.64,79.69) and (307.2,83.12) .. (302.97,83.12) -- (191.73,83.08) .. controls (187.49,83.08) and (184.06,79.64) .. (184.07,75.41) -- cycle ;
        \draw [color={rgb, 255:red, 245; green, 166; blue, 35 }  ,draw opacity=1 ]   (288.44,44.67) -- (288.88,28.7) ;
        \draw [shift={(288.93,26.7)}, rotate = 91.57] [color={rgb, 255:red, 245; green, 166; blue, 35 }  ,draw opacity=1 ][line width=0.75]    (10.93,-3.29) .. controls (6.95,-1.4) and (3.31,-0.3) .. (0,0) .. controls (3.31,0.3) and (6.95,1.4) .. (10.93,3.29)   ;
        \draw [color={rgb, 255:red, 208; green, 2; blue, 27 }  ,draw opacity=1 ]   (377.8,43.46) -- (377.43,27.95) ;
        \draw [shift={(377.38,25.95)}, rotate = 88.61] [color={rgb, 255:red, 208; green, 2; blue, 27 }  ,draw opacity=1 ][line width=0.75]    (10.93,-3.29) .. controls (6.95,-1.4) and (3.31,-0.3) .. (0,0) .. controls (3.31,0.3) and (6.95,1.4) .. (10.93,3.29)   ;
        
        \draw (248.98,108.48) node [anchor=north west][inner sep=0.75pt]  [rotate=-0.04] [align=left] {$ $};
        \draw (326.19,116) node [anchor=north west][inner sep=0.75pt]  [font=\small] [align=left] {$\displaystyle {U}$};
        \draw (246.07,116) node [anchor=north west][inner sep=0.75pt]  [font=\small] [align=left] {$\displaystyle {C}$};
        \draw (271.6,69.8) node [anchor=north west][inner sep=0.75pt]  [font=\small] [align=left] {$\displaystyle \hat{2}$};
        \draw (195.48,69.8) node [anchor=north west][inner sep=0.75pt]  [font=\small] [align=left] {$\displaystyle \hat{3}$};
        \draw (346.59,69.8) node [anchor=north west][inner sep=0.75pt]  [font=\small] [align=left] {$\displaystyle \hat{1}$};
        \draw (133.79,69.8) node [anchor=north west][inner sep=0.75pt]  [font=\small] [align=left] {$\displaystyle \hat{4}$};
        \draw (242.91,133.61) node [anchor=north west][inner sep=0.75pt]   [align=left] {$\displaystyle x_{c}( t^*_{f}$), $\displaystyle v_{c}( t^*_{f}$)};
        \draw (371.97,151.6) node [anchor=north west][inner sep=0.75pt]  [font=\small,color={rgb, 255:red, 128; green, 128; blue, 128 }  ,opacity=1 ] [align=left] {$\displaystyle L_{f}$};
        \draw (174.46,151.65) node [anchor=north west][inner sep=0.75pt]  [font=\small,color={rgb, 255:red, 128; green, 128; blue, 128 }  ,opacity=1 ] [align=left] {$\displaystyle L_{r}$};
        \draw (355.74,11.0) node [anchor=north west][inner sep=0.75pt]  [color={rgb, 255:red, 208; green, 2; blue, 27 }  ,opacity=1 ] [align=left] {$\displaystyle S(t^*_f)$};
        \draw (250.86,11.0) node [anchor=north west][inner sep=0.75pt]  [color={rgb, 255:red, 245; green, 166; blue, 35 }  ,opacity=1 ] [align=left] {$\displaystyle \left(i^{*},i^*+1\right)$};
    \end{tikzpicture}
    \caption{\centering{CAV set $S(t^*_f)$ and optimal CAV subset $(i^*,i^*+1) \in S(t^*_f)$ selection diagram}}
    \label{fig3:cav_set_selection}
    \vspace*{-\baselineskip}
\end{figure}

Let $C$ be the vehicle that initiates an automated maneuver. This can be manually triggered by the driver of $C$ deciding to overtake vehicle $U$
or automatically triggered by a given distance detected from an uncontrollable vehicle $U$ ahead of $C$, as shown in Fig. \ref{fig3:cav_set_selection}. Assuming that all vehicles other than $U$ are CAVs, we will henceforth refer to them as such.

Let $S(t)$ be a set of vehicles on the left (fast) lane which are in the vicinity of $C$ at time $t$ and contains all candidate vehicles to cooperate with $C$ in planning its lane-changing maneuver. For simplicity, once $N$ members of $S(t)$ are fixed, their indices are ordered $\{1,\ldots,N\}$ starting with the CAV furthest ahead of $C$ so that $i+1$ denotes the CAV immediately following $i$ (see Fig. \ref{fig3:cav_set_selection}). This set is limited by the communication range between $C$ and other vehicles in its vicinity, but it may otherwise be selected based on criteria such as limiting the number of candidate vehicles so as to ensure a computationally feasible solution of the cooperative pair selection.
As shown in  Fig. \ref{fig3:cav_set_selection}, we use the parameters $L_r$ and $L_f$ to define this set, where $L_r$ is a given backward distance from the rear end of $C$ and $L_f$ is a forward distance from the front of vehicle $U$. 
Thus, letting $x_i(t)$ denote the longitudinal position of vehicle $i$ along its current lane with respect to a given origin $O$, we define $S(t)$ to consist of fast lane CAVs as follows:
\begin{equation}
    \label{eq8:subset_definition}
    \begin{aligned}
        S(t) \coloneqq & \left\{i \;\; | \;\; x_i(t)\leq x_U(t)+L_f,  \right. \\
        & \: x_i(t)\geq x_C(t)-L_r \}    
    \end{aligned}
\end{equation}
It is now clear that any pair of vehicles in $S(t)$ selected to cooperate with $C$ a time $t$ is of the form $(i,i+1)$ with $i,i+1 \in S(t)$. An optimal pair, selected as described in the sequel, is, therefore, a subset of $S(t)$ denoted by $S^*(t) = \{i^*,i^*+1\}$. If multiple maneuvers are to be executed, we index them by $k=1,2,\ldots$ and write $S_k(t)$ to represent the associated set corresponding to the specific CAV $C$ that initiates the maneuver. Since we initially focus on a particular maneuver for a given $C$, we first limit ourselves to the simpler notation $S(t)$.

For every vehicle $i\in S(t)$ its dynamics take the form
\begin{equation}
    \left[ \begin{matrix}
         \dot{x}_i(t)
         \\ 
         \dot{v}_i(t)
        
    \end{matrix} \right] =
    \left[ \begin{matrix}
     v_i(t)
     \\ 
     u_i(t)
    \end{matrix} \right],
    \label{eq:vehicle_dynamics}
\end{equation}
where, in addition to $x_i(t)$, we define $v_i(t)$ and $u_i(t)$ to be vehicle $i$'s velocity and (controllable) acceleration respectively.  Without loss of generality, we define the origin for CAV $i$ involved in a maneuver to be the position $x_C(t_{0})$ of CAV $C$, where $t_{0}$ denotes the time at which the maneuver starts. We will use $t_f$ to denote the time when the longitudinal maneuver is complete.

The control input and speed are constrained as follows:
\begin{equation}
    \begin{matrix}
        u_{i_{\min}}\leq u_i(t)\leq u_{i_{\max}}, \; \; \forall t\in\lbrack t_{0},t_{f}\rbrack
        \\
        v_{i_{\min}}\leq v_i(t)\leq v_{i_{\max}}, \; \; \forall t\in\lbrack t_{0},t_{f}\rbrack
    \end{matrix},
    \label{eq:vehicle_constraints}
\end{equation}
where $v_{i_{\max}}>0$ and $v_{i_{\min}}>0$ denote the maximum and minimum speed allowed, usually determined by the rules of the highway and they may be common for all $i\in S(t)$. Similarly, $u_{i_{\max}}>0$ and $u_{i_{\min}}<0$ represent vehicle $i$'s maximum and minimum acceleration control.

{\bf Safety Constraints}. Let $d_i(v_i(t))$ be the speed-dependent safety distance of CAV $i$, defined as the minimum required distance between $i$ and its immediately preceding vehicle traveling on the same lane. Specifically, we define:
\begin{equation}
    d_i(v_i(t)) = \varphi  v_i(t)+\delta,
    \label{eq:safety_constraint}
\end{equation}
where $\varphi$ denotes the reaction time (usually set to $\varphi =1.8\,s$ \cite{vogel2003comparison}). The safety distance $d_i(v_i(t))$ is specified from the center of vehicle $i$ to the center of its preceding vehicle. Thus, a $\delta$ constant value is selected to capture at least the vehicle dimensions augmented by any desired value (e.g. $1.5\,m$).

We can now define all safety constraints that must be satisfied during a lane-changing maneuver of $C$ when cooperating with any two CAVs $(i,i+1)$:
\begin{subequations}
    \begin{align}
        x_{U}(t)-x_{C}(t)&\geq d_{C}(v_{C}(t)),\; \; \; \; \, \forall  t\in\lbrack t_0,t_{f}] \label{eq4a:cavC_vehU_safety_constraint}\\ 
        x_{i}(t)-x_{i+1}(t)&\geq d_{i+1}(v_{i+1}(t)), \; \; \;\forall t\in\lbrack t_0,t_{f}] \label{eq4b:cavi_cavi1_safety_constraint}\\ 
        x_{C}(t_{f})-x_{i+1}(t_{f})&\geq d_{i+1}(v_{i+1}(t_{f})), \label{eq4c:cavC_cavi1_safety_constraint}\\
        x_{i}(t_{f})-x_{C}(t_{f})&\geq d_{C}(v_{C}(t_{f})) \label{eq4d:cavi_cavC_safety_constraint}
     \end{align}
    \label{eq:safety_constraint}
    \vspace*{-\baselineskip}
\end{subequations}

{\bf Traffic Disruption}. We now seek to measure the extent to which a successful lane-changing maneuver may disrupt the left-lane traffic. 
Given any time $t_f>t_0$ and $T=t_f-t_0$, let $x_i(t_f)$ be the terminal position of $i$ as determined by some control policy $u_i(t)$, $t\in \lbrack t_0,t_f]$. We then define 
\begin{equation}
    \Delta_{i}(T) = x_i(t_f)-\lbrack x_i(t_0)+v_i(t_0)T \rbrack
    \label{eq:Ideal}
\end{equation}
which specifies the difference between the actual  terminal position of $i$ under some control policy and its ideal terminal position obtained by maintaining a constant speed $v_i(t_0)$. This is ``ideal'' in the sense that the vehicle's uniform motion is undisrupted, hence also minimizing the energy consumption which would be due to any acceleration/deceleration. This motivates the definition of the following maneuver disruption metric $D_{i,i+1}(T)$ applied to two cooperating CAVs $i,i+1$:
\begin{equation}
    D_{i,i+1}(T) = \gamma \Delta_{i}^2(T)+ \left(1-\gamma \right)\Delta^2_{i+1}(T),
    \label{eq:disruption}
\end{equation}
where $\Delta_{i}(T)$ defined in \eqref{eq:Ideal} is the disruption caused to $i$ due to an acceleration/deceleration control $u_i(t)$ applied to it relative to its undisrupted final position. The weight $\gamma \in [0,1]$ is included to potentially place more emphasis on one cooperating CAV over the other. Note that this quadratic disruption metric depends only on the total maneuver time $T$ and the terminal positions $x_i(t_f)$ and $x_{i+1}(t_f)$ for CAVs $i$ and $i+1$. One can see that $D_{i,i+1}(T)$ implicitly penalizes the time that CAV $i+1$ (which normally decelerates to accommodate $C$) would take to accelerate back to the free flow speed. Further, if $\Delta_{i+1}(T)$ is large and the vehicle following $i+1$ is closely behind, then $D_{i,i+1}(T)$ implicitly penalizes the deceleration of this vehicle and the time that it would take to accelerate back to the free flow speed. 
Clearly, if there are multiple maneuvers indexed by $k=1,2,\ldots$, we can define an aggregate metric
$D_\text{Total}$ by adding all individual disruptions $D^k_{i,i+1}(T_k)$ and seeking to minimize $D_\text{Total}$ by determining optimal maneuver times and terminal positions in \eqref{eq:disruption}.

{\bf Optimization Problem}. We consider two objectives for the longitudinal maneuver problem: first, we wish to minimize the maneuver time $t_f$ experienced by CAV $C$ and its cooperating vehicles; second, we wish to minimize the energy consumption of each of the three cooperating CAVs $C$, $i$ and $i+1$. At the same time, we must satisfy the safety constraints \eqref{eq:safety_constraint} and vehicle constraints \eqref{eq:vehicle_constraints}. Finally, we must ensure that the disruption metric \eqref{eq:disruption} does not exceed a given threshold $D_{th}$. Note that an alternative formulation is to seek the minimization of \eqref{eq:disruption} while keeping the maneuver time below a given threshold $T_{th}$ (this is the subject of ongoing work so as to ultimately provide comparisons between the two approaches).

The overall optimization problem is outlined next: 

1. CAV $C$ determines an optimal terminal time $t_f^*$ for the maneuver and  control $\{ u_C^*(t) \}$, $t \in [t_0,t_f^*]$ so as to minimize a given objective function $J_C$ (to be defined in the sequel) subject to the vehicle dynamics (\ref{eq:vehicle_dynamics}) and constraints 
(\ref{eq4a:cavC_vehU_safety_constraint}), (\ref{eq4c:cavC_cavi1_safety_constraint}), (\ref{eq4d:cavi_cavC_safety_constraint})
and (\ref{eq:vehicle_constraints}). Moreover, its optimal terminal speed $v_C(t_f^*)$ must be close to (or exactly match) a desired speed $v_d$ that matches the fast-lane speed. 

2. The solution $t_f^*$ specifies $S(t_f^*)$, the set from which an optimal pair $(i^*,i^*+1)$ of cooperating CAVs must be selected so as to minimize the disruption metric $D_{i,i+1}(T^*)$ in (\ref{eq:disruption}), where $T^*=t_f^*-t_0$. Since $D_{i,i+1}(T^*)$ depends on the values of $x_i(t_f^*)$ and $x_{i+1}(t_f^*)$ in (\ref{eq:Ideal}), its minimization depends on the optimal trajectories selected by CAVs $i \in S(t_f^*)$. This requires the determination of optimal controls $\{ u_i^*(t) \}$, $t \in [t_0,t_f^*]$ for all $i \in S(t_f^*)$
minimizing a given objective function $J_i$ (to be defined in the sequel) subject to the vehicle dynamics (\ref{eq:vehicle_dynamics}) and constraints
(\ref{eq4b:cavi_cavi1_safety_constraint}), (\ref{eq4c:cavC_cavi1_safety_constraint}), (\ref{eq4d:cavi_cavC_safety_constraint})
and (\ref{eq:vehicle_constraints}). 

3. Finally, we determine an optimal pair $(i^*,i^*+1)$ which minimizes the 
disruption metric $D_{i,i+1}(T^*)$ over all $i \in S(t_f^*)$. 
In addition, this solution must satisfy the requirement $D_{i^*,i^*+1}(T^*) \le D_{th}$. 

Clearly, a solution to this problem, consisting of $(i^*,i^*+1)$ and $\{ u_C^*(t),u_{i^*}^*,u_{i^*+1}^* \}$, $t \in [t_0,t_f^*]$, may not exist. For example, there is no guarantee that a pair $(i^*,i^*+1)$ can be found satisfying the disruption requirement $D_{i^*,i^*+1}(T^*) \le D_{th}$. 

In the next section, we present a detailed solution approach with the following key elements: (i) We specify the objective functions $J_C$ and $J_i$ for any $i \in S(t)$. 
(ii) We obtain the optimal cooperating pair $(i^*,i^*+1)$ without solving all (time-consuming) optimal control problems for $\{ u_i^*(t) \}$, $t \in [t_0,t_f^*]$.
(iii) We include a time relaxation on $t_f^*$ so that if a feasible solution does not exist, we seek one for a relaxed value $t'_f > t_f^*$. Intuitively, it may not be possible to achieve an acceptable disruption below $D_{th}$ under a short maneuver time $t_f^*$ which is ``selfishly'' set by $C$; thus, this relaxation process captures the trade-off between the ``selfish'' goal of $C$ to minimize its maneuver time and the system-wide ``social'' goal of minimizing traffic flow disruptions and ensuring an increase in throughput by the addition of vehicles to the fast lane.

%% file: sections/optimal_control_solution.tex
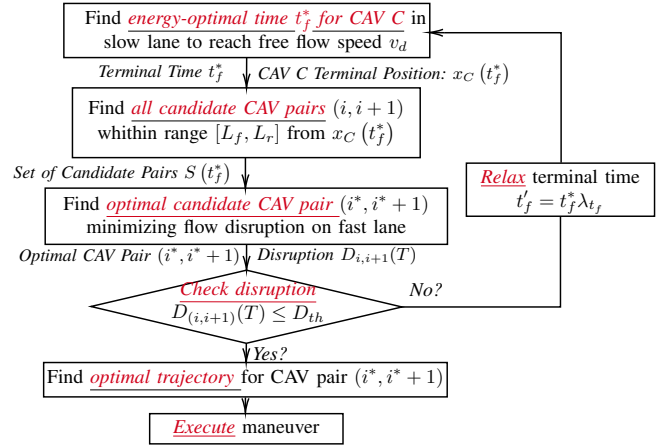
\begin{figure}[pt]
    \centering
    \vspace*{\baselineskip}
    \tikzset{every picture/.style={line width=0.75pt}} 
    \resizebox{\textwidth/2-6pt}{!}{
        \begin{tikzpicture}[x=0.66pt,y=0.53pt,yscale=-1,xscale=1]
        
        \draw   (278.07,257.85) -- (397.82,293.19) -- (277.91,328.07) -- (158.15,292.73) -- cycle ;
        \draw    (278.1,233.08) -- (278.06,255.83) ;
        \draw [shift={(278.06,257.83)}, rotate = 270.1] [color={rgb, 255:red, 0; green, 0; blue, 0 }  ][line width=0.75]    (10.93,-3.29) .. controls (6.95,-1.4) and (3.31,-0.3) .. (0,0) .. controls (3.31,0.3) and (6.95,1.4) .. (10.93,3.29)   ;

        \draw    (277.91,328.17) -- (277.82,344.55) ;
        \draw [shift={(277.81,346.55)}, rotate = 270.29] [color={rgb, 255:red, 0; green, 0; blue, 0 }  ][line width=0.75]    (10.93,-3.29) .. controls (6.95,-1.4) and (3.31,-0.3) .. (0,0) .. controls (3.31,0.3) and (6.95,1.4) .. (10.93,3.29)   ;
        \draw    (430.35,29.79) -- (420.5,29.62) ;
        \draw [shift={(418.5,29.58)}, rotate = 1.02] [color={rgb, 255:red, 0; green, 0; blue, 0 }  ][line width=0.75]    (10.93,-3.29) .. controls (6.95,-1.4) and (3.31,-0.3) .. (0,0) .. controls (3.31,0.3) and (6.95,1.4) .. (10.93,3.29)   ;
        \draw   (430.35,29.79) -- (520.5,29.79) -- (520.5,154.58) ;
        
        \draw   (397.82,293.19) -- (520,293.19) -- (520,206.05) ;
        \draw    (277.21,380.08) -- (277.21,392.58) ;
        \draw [shift={(277.21,394.58)}, rotate = 270] [color={rgb, 255:red, 0; green, 0; blue, 0 }  ][line width=0.75]    (10.93,-3.29) .. controls (6.95,-1.4) and (3.31,-0.3) .. (0,0) .. controls (3.31,0.3) and (6.95,1.4) .. (10.93,3.29)   ;
        \draw    (277.21,149.35) -- (277.21,177.58) ;
        \draw [shift={(277.21,179.58)}, rotate = 270] [color={rgb, 255:red, 0; green, 0; blue, 0 }  ][line width=0.75]    (10.93,-3.29) .. controls (6.95,-1.4) and (3.31,-0.3) .. (0,0) .. controls (3.31,0.3) and (6.95,1.4) .. (10.93,3.29)   ;
        
        \draw (212.5,68.39) node  [font=\small] [align=left] {\textit{Terminal Time }$\displaystyle t_{f}^{*}$};
        \draw (384.21,69.55) node  [font=\small] [align=left] {\textit{CAV C Terminal Position: }$\displaystyle x_{C}\left( t_{f}^{*}\right)$};
        \draw (184.91,165.35) node  [font=\small] [align=left] {\textit{Set of Candidate Pairs }$\displaystyle S\left( t_{f}^{*}\right) \ $};
        \draw    (203.41,395.66) -- (352.41,395.66) -- (352.41,424.66) -- (203.41,424.66) -- cycle  ;
        \draw (277.91,410.16) node   [align=left] {\begin{minipage}[lt]{98.49pt}\setlength\topsep{0pt}
        \begin{center}
        \textcolor[rgb]{0.82,0.01,0.11}{\textit{\underline{Execute}}} maneuver
        \end{center}
        
        \end{minipage}};
        \draw (279.2,289.97) node  [rotate=-0.11,xslant=0] [align=left] {\begin{minipage}[lt]{85.34pt}\setlength\topsep{0pt}
        \begin{center}
        \textit{\textcolor[rgb]{0.82,0.01,0.11}{\underline{Check disruption}}} \\$\displaystyle D_{( i,i+1)}(T) \leq D_{th}$
        \end{center}
        
        \end{minipage}};
        \draw (280.91,329.83) node [anchor=north west][inner sep=0.75pt]   [align=left] {\textit{Yes?}};
        \draw (402.26,268.33) node [anchor=north west][inner sep=0.75pt]   [align=left] {\textit{No?}};
        \draw    (143.91,82.75) -- (411.91,82.75) -- (411.91,149.75) -- (143.91,149.75) -- cycle  ;
        \draw (277.91,116.25) node   [align=left] {\begin{minipage}[lt]{179.52pt}\setlength\topsep{0pt}
        \begin{center}
        Find \underline{\textcolor[rgb]{0.82,0.01,0.11}{\textit{all candidate CAV pairs}}} $\displaystyle ( i,i+1)$ whithin range $\displaystyle [ L_{f} ,L_{r}]$ from $\displaystyle x_{C}\left( t_{f}^{*}\right)$
        \end{center}
        
        \end{minipage}};
        \draw    (137.29,1.67) -- (419.29,1.67) -- (419.29,53.67) -- (137.29,53.67) -- cycle  ;
        \draw (140.29,5.67) node [anchor=north west][inner sep=0.75pt]   [align=left] {\begin{minipage}[lt]{189.01pt}\setlength\topsep{0pt}
        \begin{center}
        Find \textit{\underline{\textcolor[rgb]{0.82,0.01,0.11}{energy-optimal time }}}\textcolor[rgb]{0.82,0.01,0.11}{$\displaystyle t_{f}^{*}$}\textit{\underline{\textcolor[rgb]{0.82,0.01,0.11}{\ for CAV C}}} in \\slow lane to reach free flow speed $\displaystyle v_{d}$
        \end{center}
        
        \end{minipage}};
        \draw    (447.5,155.05) -- (592.5,155.05) -- (592.5,206.05) -- (447.5,206.05) -- cycle  ;
        \draw (520,159.05) node [anchor=north] [inner sep=0.75pt]   [align=left] {\begin{minipage}[lt]{95.9pt}\setlength\topsep{0pt}
        \begin{center}
        \textcolor[rgb]{0.82,0.01,0.11}{\textit{\underline{Relax}}} terminal time \\$\displaystyle t_{f}' =t_{f}^{*}\lambda_{t_f}$
        \end{center}
        
        \end{minipage}};
        \draw (187.91,244.68) node  [font=\small] [align=left] {\begin{minipage}[lt]{121.31pt}\setlength\topsep{0pt}
        \begin{center}
        \textit{Optimal CAV Pair }$\displaystyle \left( i^{*} ,i^{*} +1\right)$
        \end{center}
        
        \end{minipage}};
        \draw    (122.91,179.05) -- (432.91,179.05) -- (432.91,233.05) -- (122.91,233.05) -- cycle  ;
        \draw (277.91,206.05) node   [align=left] {\begin{minipage}[lt]{207.89pt}\setlength\topsep{0pt}
        \begin{center}
        Find \textcolor[rgb]{0.82,0.01,0.11}{\textit{\underline{optimal candidate CAV pair }}}$\displaystyle \left( i^{*} ,i^{*} +1\right)$ minimizing flow disruption on fast lane
        \end{center}
        
        \end{minipage}};
        \draw (347.91,243.68) node  [font=\small] [align=left] {\begin{minipage}[lt]{83.22pt}\setlength\topsep{0pt}
        \begin{center}
        \textit{Disruption }$\displaystyle D_{i,i+1}( T)$
        \end{center}
        
        \end{minipage}};
        \draw    (121.71,346.55) -- (432.71,346.55) -- (432.71,379.55) -- (121.71,379.55) -- cycle  ;
        \draw (277.21,363.05) node   [align=left] {\begin{minipage}[lt]{208.57pt}\setlength\topsep{0pt}
        \begin{center}
        Find \textit{\underline{\textcolor[rgb]{0.82,0.01,0.11}{optimal trajectory }}\textcolor[rgb]{0,0,0}{ }}\textcolor[rgb]{0,0,0} {for CAV pair }\textcolor[rgb]{0,0,0}{$\displaystyle ( i^*,i^*+1)$}
        \end{center}
        
        \end{minipage}};
        \draw    (278.17,53.67) -- (278.06,80.75) ;
        \draw [shift={(278.05,82.75)}, rotate = 270.24] [color={rgb, 255:red, 0; green, 0; blue, 0 }  ][line width=0.75]    (10.93,-3.29) .. controls (6.95,-1.4) and (3.31,-0.3) .. (0,0) .. controls (3.31,0.3) and (6.95,1.4) .. (10.93,3.29)   ;
        \end{tikzpicture}
    }
    \caption{Cooperative Maneuver Flow Diagram  }
    \label{Fig2:MinimalDisruptionFlowchart}
    \vspace*{-\baselineskip}
\end{figure}

The overall description of the solution process for the optimal maneuver ensuring a minimal disruption that satisfies a given threshold $D_{th}$ is given in Fig. \ref{Fig2:MinimalDisruptionFlowchart}. A step-by-step detailed explanation of this process is given below and is also shown in Alg. \ref{alg1:Optimal_longitudinal_lane-change}.
\subsection{CAV C Optimal Trajectory} \label{SubSec3_A:CAVC_Optimal_Trajectory}
Given a CAV $C$ traveling behind an uncontrolled vehicle $U$, a ``start maneuver'' request is sent to surrounding vehicles and the starting time $t_0$ is defined when $x_U(t)-x_C(t)\leq d_\text{start}$, where $d_\text{start}$ denotes the minimum distance at which CAV $C$  decides to initiate a lane-changing maneuver. It is important to point out that at that instant the values of the optimal maneuver time and the entire optimal trajectory of $C$ can be evaluated, which also enables planning the complete solution of the problem.

We now formulate the optimal control problem (OCP) that $C$ solves by first defining its objective function:

\begin{equation} 
        J_{C}(t_f,u_C(t)) = 
        \int_{t_0}^{t_f}\left(\alpha+\dfrac{\left(1-\alpha\right)u^2_C(t)}{\max\left\{ {u_{C_{\min}}^2}, {u_{C_{\max}}^2}\right\}}\right)dt
    \label{eq:cav_c_objective}
\end{equation}
where $\alpha \in \lbrack0,1\rbrack$ is an adjustable weight that penalizes travel time relative to the total energy cost for CAV $C$. Note that by properly normalizing the energy cost term $u^2_C(t)$, the integrand above is a convex combination of time and an energy metric. 
Letting $\beta \coloneqq \dfrac{\alpha\max\left\{ {u_{C_{\min}}^2}, {u_{C_{\max}}^2}\right\}}{2\left(1-\alpha\right)}$, the following notationally simpler optimal control problem formulation is obtained:
\begin{equation} 
    \min _{\{t_f,u_{C}(t)\}} \beta \left(t_f-t_0\right)+
    \int_{t_0}^{t_f}\dfrac{1}{2}{u_C}^2(t)dt
    \label{eq:cav_c_objective_simplified}
    \vspace*{-\baselineskip}
\end{equation}
\begin{gather*}
    \begin{matrix}
        \text{s.t. \ \eqref{eq:vehicle_dynamics}, \eqref{eq:vehicle_constraints},}\\
        x_{U}(t)-x_{C}(t) \geq d_{C}(v_{C}(t)), && \forall  t\in\lbrack t_0,t_{f}]\\
        \left(v_{C}(t_f)-v_d\right)^2 \leq {\delta_\text{tol}},\\
        0\leq t_f \leq T_{th}
     \end{matrix}
\end{gather*} 
where the last three constraints capture (i) the safe distance constraint \eqref{eq4a:cavC_vehU_safety_constraint} between $C$ and $U$ (assuming that the position and speed of $U$ can be sensed or estimated by $C$), 
(ii) the requirement that the terminal speed $v_{C}(t_f)$ matches a desired speed $v_d$ within a tolerance $\delta_{\text{tol}} \ge 0$ and 
(iii) $T_{th}$ is specified as the maximum tolerable time to perform a lane-changing maneuver. In practice, if the last constraint cannot be met for a given $T_{th}$, CAV $C$ has the option of either relaxing this value or simply aborting the maneuver. Finally, note that problem (\ref{eq:cav_c_objective_simplified}) is solved given the initial position and speed of CAV $C$.

The solution of this OCP can be analytically obtained through standard Hamiltonian analysis similar to OCPs formulated and solved in \cite{chen2020cooperative}, therefore, we omit the details. It is worth pointing out, however, that depending on the weight $\beta$ and the starting distance $d_\text{start}$ of the maneuver, the form of the corresponding optimal trajectory can be either strictly \emph{accelerating} or first \emph{decelerating} followed by an accelerating component so that \eqref{eq4a:cavC_vehU_safety_constraint} is satisfied regardless of the initial conditions of CAV C. Intuitively, if there is adequate distance ahead of $C$, it can accelerate at a maximal rate to attain $v_c(t_f) = v_d$; otherwise, it needs to first decelerate to create such an adequate distance ahead of it and then accelerate to meet the terminal speed constraint in (\ref{eq:cav_c_objective_simplified}).
\subsection{Construction of Cooperative Set $S(t_f^*)$} \label{SubSec3_B:optimal_collaboration_pair_selection}
Once \eqref{eq:cav_c_objective_simplified} is solved, the optimal terminal time $t_f^*$ and terminal position $x_C(t_f^*)$ are determined.
Therefore, the cooperative set $S(t_f^*)$ can be determined from (\ref{eq8:subset_definition}) over different feasible positions 
$x_i(t_f^*)$ of CAVs $i$.
In particular, we can determine optimal trajectories for each $i \in S(t_0)$ based on an OCP solved by $i$ to minimize its energy and then use $x_i^*(t_f^*)$ in $S(t_f^*)$.
However, we exploit the fact that the actual set $S(t_f^*)$ of candidate CAVs is unaffected as long as the conditions 
$x_i(t)\leq x_U(t)+L_f$ and $x_i(t)\geq x_C(t)-L_r$ are satisfied. Thus, rather than the time-consuming process of solving OCPs for all $i$,
we define $S(t_f^*)$ as follows:
\begin{equation}
    \label{eq9:subset_definition}
    \begin{aligned}
        S(t_f^*) \coloneqq & \left\{i \;\; | \;\; x_{i}(t_0)+v_{i}(t_f^*- t_0) \leq x_U(t_f^*)+L_f,  \right. \\
        & \: x_{i}(t_0)+v_{i}(t_f^*- t_0)\geq x_C(t_f^*)-L_r \},    
    \end{aligned}
\end{equation}
where  $x_{i}(t_0)+v_{i}(t_f^*- t_0)$ is the position of $i$ at time $t_f^*$ under constant speed. Alternatively, we can use in (\ref{eq8:subset_definition}) the value of $x_i(t_f^*)$ under maximal acceleration to determine a set $S_{\max}(t_f^*)$ and maximal deceleration 
to determine a set $S_{\min}(t_f^*)$ and then define $S(t_f^*) = S_{\max}(t_f^*) \cup S_{\min}(t_f^*)$. Also note that by adjusting the parameters $L_f$ and $L_r$ the size of this set can be adjusted to include as many candidate CAVs as desired, subject only to the constraint that any $S(t)$ must include CAVs within the communication range of CAV $C$ to allow full cooperation.

\subsection{Optimal Cooperative Pair $(i^*,i^*+1)$}\label{SubSec3B3:optimalSubset}
\begin{figure}[hpt]
    \vspace*{-\baselineskip}
    \includegraphics[height=5.5cm,width = 9cm-4pt]{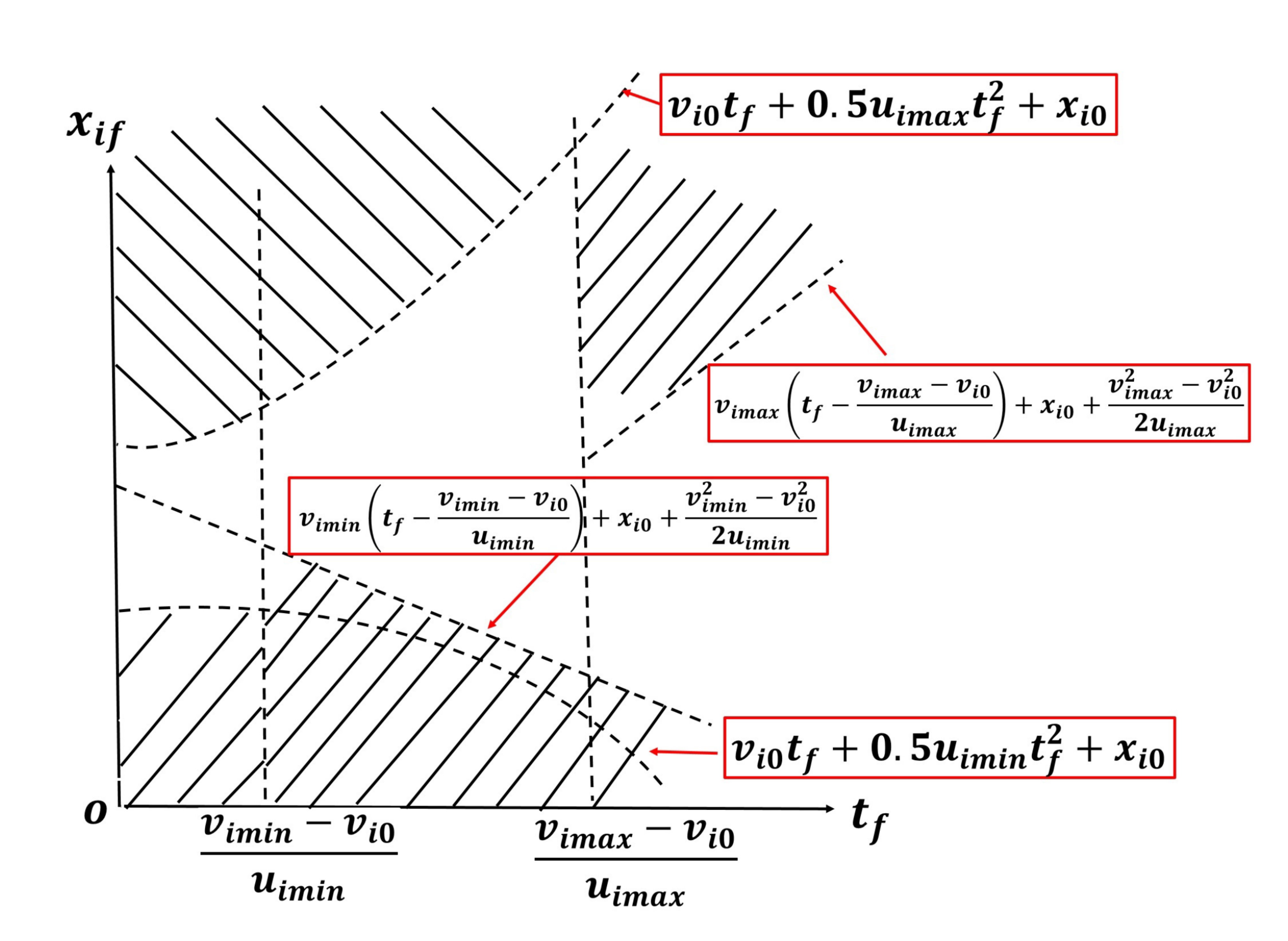}
    \caption{Feasible Terminal Set $X_{i}^{feas}(t_f)$ Visualization}
    \label{fig4:feasible_set}

    \vspace*{-\baselineskip}
\end{figure}

The optimal cooperative pair $(i^*,i^*+1)$ among all $i \in S(t_f^*)$ is the one that minimizes the disruption metric in (\ref{eq:disruption}) by selecting terminal positions resulting in minimal disruption.
These terminal positions $x_i(t_f^*),x_{i+1}(t_f^*)$ are subject to the vehicle constraints specified in \eqref{eq:vehicle_constraints}. As a result, a set of feasible terminal positions can be defined as shown in Fig. 2, Section III-A of \cite{chen2020cooperative}. For convenience, this is reproduced here as Fig. \ref{fig4:feasible_set}. Therefore, each $x_i(t_f^*)$
must be constrained to this set which we denote by $X_{i}^{feas}(t_f^*)$.

We can now formulate the following Quadratic Program (QP) whose solution provides the optimal terminal positions $x_i(t_f^*),x_{i+1}(t_f^*)$ for any pair $i,i+1 \in S(t_f^*)$:
\begin{equation}
    \begin{aligned}
        &\min_{x_{i}(t_f^*),x_{i+1}(t_f^*)}
        \gamma\Delta^2_{i}(T)+
        \left(1-\gamma \right) \Delta^2_{i+1}(T)\\
        \text{s.t. \ }
        &x_{i}(t_f^*)-x_C(t_f^*)\geq d_C\left(v_C(t_f^*)\right),\\
        &x_C(t_f^*)-x_{i+1}(t_f^*)\geq \max \left\{d_{i+1}\left(v_{i+1}(t_f^*)\right)\right\},\\
        &x_{i-1}(t_f^*)-x_{i}(t_f^*)\geq \max \left\{d_{i}\left(v_{i}(t_f^*)\right)\right\},\\
        &x_i(t_f^*) \in X_{i}^{feas}(t_f^*), \;\; x_{i+1}(t_f^*) \in X_{i+1}^{feas}(t_f^*)\\
        &T=t_f^*-t_0
    \end{aligned}
    \label{eq:terminal_position}
\end{equation}
where the first three constraints are the safe distance requirements in (\ref{eq:safety_constraint}) that $i$ and $i+1$ must satisfy.
The values of the $\max \left\{\cdot\right\}$ terms in \eqref{eq:terminal_position} are assumed to be given by prespecified maximum safe distances;  for simplicity, $\max \{d_{i+1}(v_{i+1}(t_f^*))\}=d_{i+1}(v_{i+1}(t_0))$ and $\max \{d_{i}(v_{i+1}(t_f^*))\}=d_{i}(v_{i}(t_0)+{u_{i_{\max}}}( t_f^*-t_0))$. 
Note that in the third constraint it is possible that $i-1=0$, i.e., CAV $i$ is the first in the set $S(t_f^*)$ and there may not always exist a vehicle ahead of it, in which case we simply assign $x_{i-1}(t_f^*)$ an arbitrarily large value (e.g., $x_{i-1}(t_f^*)=x_{i}(t_f^*)+1000m$).

The solution of \eqref{eq:terminal_position} for every pair $(i,i+1)$ provides the optimal terminal positions $x_i^*(t_f^*)$ and $x_{i+1}^*(t_f^*)$ that minimize the disruption metric $D_{i,i+1}(T)$. We shall denote this optimal value as $D^*_{i,i+1}(T)$. 
We can now determine the optimal cooperative pair $(i^*,i^*+1)$ from
\begin{equation}
    \begin{aligned}
        (i^*,i^*+1) &= \argmin_{(i,i+1) \in S(t_f^*)} D^*_{i,i+1}(T)\\
        \text{s.t.} &\;\; D^*_{i,i+1}(T) \leq D_{th}
    \end{aligned}
    \label{eq:optimal_collaboration_pair}
\end{equation}
This is a simple minimization problem comparing the values of $D^*_{i,i+1}(T)$ obtained from (\ref{eq:terminal_position}) over a finite set consisting of pairs $(i,i+1)$ that satisfy the disruption constraint above.

If no solution to (\ref{eq:optimal_collaboration_pair}) is found, it is still possible to derive a solution based on the analysis in \cite{chen2020cooperative}, with no consideration of disruption. 
Alternatively, we may proceed with the time relaxation process described in Section \ref{SubSec3_C:Time_Relaxation}. However, we first complete the solution process by computing optimal trajectories for CAVs $(i^*,i^*+1)$ if such a pair is identified.

\subsection{Optimal Trajectories for CAVs $(i^*,i^*+1)$} \label{SubSec_A:CAV_i_i+1_Optimal_Trajectories}
Assuming an optimal cooperative pair $(i^*,i^*+1)$ has been determined that satisfies all the problem constraints, it remains to specify optimal trajectories for these two CAVs.
For any such CAV $i$, we define its objective function to be:
\begin{equation} 
        J_{i}(u_i(t)) = 
        \int_{t_0}^{t_f^*} \frac{1}{2} u_i^2(t)dt
    \label{eq:cav_i_objective}
\end{equation}
so that CAV $i$ (where $i=i^*$ or $i=i^*+1$) solves the following fixed terminal time and position OCP:
\begin{equation} 
    \min _{\{u_{i}(t)\}} 
    \int_{t_0}^{t_f^*}\dfrac{1}{2}{u_i}^2(t)dt
    \label{eq:cav_i_problem}
    \vspace*{-\baselineskip}
\end{equation}
\begin{gather*}
    \begin{matrix}
        \text{s.t. \ \eqref{eq:vehicle_dynamics}, \eqref{eq:vehicle_constraints}}, x_{i}(t_f^*) = x^*_i(t_f^*)
     \end{matrix}
\end{gather*} 
where $x^*_i(t_f^*)$ is the optimal terminal position in (\ref{eq:terminal_position}) when $i=i^*$ or $i=i^*+1$. This is an OCP of the same form as those solved in \cite{chen2020cooperative}, thus, we omit the details of the solution.

\subsection{Maneuver Time Relaxation} \label{SubSec3_C:Time_Relaxation}
As already mentioned, it is possible that no solution to problem \eqref{eq:optimal_collaboration_pair} may be found. The most common reason is due to the fact that the optimal maneuver end time $t_f^*$ determined by CAV $C$ at the first step of the solution approach is too short to allow $C$ to reach a speed sufficiently close to $v_d$  and for cooperating CAVs to adjust their positions so as to satisfy the safety constraints in (\ref{eq:safety_constraint}).
In such cases, it is possible to perform a relaxation of $t_f^*$  obtained through \eqref{eq:cav_c_objective_simplified} by trading it off against the energy consumption due to the maneuver extension. Thus, the new terminal time is given as $t'_{f}=t_{f}\lambda_{t_f}$ where $\lambda_{t_f}>1$ is a relaxation factor. Observe that this time modification changes the form of the OCP \eqref{eq:cav_c_objective_simplified}, since the terminal time is now fixed at $t'_{f} > t_f^*$ and the solution will lead to a new terminal position $x_c(t'_f)$ for CAV $C$. The new OCP formulation is as follows:
\begin{equation} 
    \min _{\{u_{C}(t)\}} J_{C}(u_C(t))= 
    \int_{t_0}^{{t}_f'}\dfrac{1}{2}{u_C^2(t)} dt
    \label{eq:cav_c_relaxedTime}
    \vspace*{-\baselineskip}
\end{equation}
\begin{gather*}
    \text{s.t. \ (\ref{eq:vehicle_dynamics}), (\ref{eq:vehicle_constraints}), and}\\
    \begin{aligned} 
        x_{U}(t)-x_{C}(t) &\geq d_{C}(v_{C}(t)),\text{ \ \ } \forall  t\in\lbrack t_0,t_{f}],\\
        \left(v_{C}(t_f')-v_d\right)^2 &\leq \delta_\text{tol}
    \end{aligned}
\end{gather*} 
This process may continue, as shown in Fig. \ref{Fig2:MinimalDisruptionFlowchart}, until a feasible solution is determined or the constraint $T \le T_{th}$ is violated.

\textit{\textbf{Remark}}: Despite time relaxation, problem \eqref{eq:cav_c_relaxedTime} can still be infeasible if $d_\text{start}$ is small or if the constraint \eqref{eq4a:cavC_vehU_safety_constraint}  is active at $t_0$. Therefore, CAV $C$ can abort the maneuver and wait a specified time interval for the next opportunity window. Otherwise, a ``selfish'' maneuver may be performed as in \cite{chen2020cooperative} by computing the minimum feasible terminal time and minimum terminal position for any $i$ and $i+1$ with $i\in S(t_f^*)$.

\subsection{Sequential Maneuvers} \label{SubSec3_F:Sequential_Maneuvers}
Clearly, we can perform a series of individual maneuvers following Fig. \ref{Fig2:MinimalDisruptionFlowchart} indexed by $k$ that minimize the aggregate metric $D_\text{Total}=\sum^N_{k=1}D^k_{i,i+1}(T_k)$. For each maneuver $k$ we can compute a series of system-centric (social) optimal trajectories that start sequentially by defining a CAV $C$ as soon as maneuver $k-1$ has completed its corresponding lateral phase. Thus, the initial time for maneuver $k$ is upper bounded by the terminal time of maneuver $k-1$ $\left(t^k_0\geq t^{k-1}_f\right)$. Note that CAV $C$ for maneuver $k-1$ can become a CAV candidate for the $k$th maneuver. It is also possible to \emph{parallelize} a number of such maneuvers by allowing them to start simultaneously given a set of target vehicles (CAV $C$). This is the subject of ongoing work which will allow us to evaluate the extent to which such parallelization is possible and whether it may outperform the sequential process. 

\begin{algorithm}[hpt]
    \caption{Optimal Longitudinal Maneuver}
    \SetKwInOut{Input}{input}
    \SetKwInOut{Output}{output}
    \Input{Initial Conditions $x_i(t_0),v_i(t_0)\coloneqq\{i|i\in S(t)\}$,  $x_U(t_0),x_U(t_0)$, and $x_C(t_0),x_C(t_0)$, Relaxation Constant $\lambda_{t_f}$,\\Free Flow speed $v_{d}$, Maximum Time $T_{th}$,\\ Maximum Disturbance $D_\text{th}$} 
    \Output{Optimal Longitudinal Trajectories $u_i^*(t) \coloneqq \{u_i^*(t)|i\in S(t^*_f), t\in \lbrack t_0,t^*_f\rbrack$\}}
    \SetKwBlock{Beginn}{beginn}{ende}
    \Begin{
        $(i^*,i^*+1)\leftarrow \emptyset$ \\ 
        $t^*_f,x_c(t_f^*),u_c^*\leftarrow$ Compute OCP for CAV C  \eqref{eq:cav_c_objective_simplified}\\
        \While{$i^*=\emptyset \wedge t_f\leq T_{th}$}{
            $S(t^*_f)\leftarrow$ Compute Relevant CAV Set \eqref{eq9:subset_definition}\\
            $i=1$, $N=|S(t^*_f)|$, $\Hat{D} = \emptyset$\\
            \For{$i=1$ to $N-1$}{
                $x_i(t^*_f),x_{i+1}(t^*_f)\leftarrow$ Compute Terminal Positions \eqref{eq:terminal_position}\\
                $D_{i,i+1}\leftarrow$ Compute Disruption \eqref{eq:disruption}\\
                \If{$D_{i,i+1}\leq D_\text{th}$}{
                    $\Hat{D}\leftarrow$ Append tuple $\left\langle i, D_{i,i+1} \right\rangle$\\
                } 
            }
            $(i^*,i^*+1)\leftarrow$ Compute optimal CAV Set \eqref{eq:optimal_collaboration_pair}\\
            \eIf{$(i^*,i^*+1)\neq\emptyset$}{
                $u_i^*,u_{i+1}^*\leftarrow$ Compute optimal trajectories for $i^*$ and $i^*+1$ \eqref{eq:cav_i_problem}\\
                break
            } 
            {
                $t'_f\leftarrow$ Relax Terminal Time $t^*_{f}\lambda_{t_f}$\\
                $x_c(t_f'),u_c^*\leftarrow$ Relaxed OCP  \eqref{eq:cav_c_relaxedTime}\\
                $\text{count}+1$
            }
        }
    }
    \label{alg1:Optimal_longitudinal_lane-change}
\end{algorithm}



%% file: sections/simulation_results.tex
In this section, we provide simulation results illustrating the time and energy-optimal controllers we have derived and comparing their performance against a baseline of non-cooperating vehicles. We test our algorithm using the traffic simulation software PTV Vissim.

\begin{figure}[ht]
    \centering
    \includegraphics[height=2cm, width = 9cm-4pt]{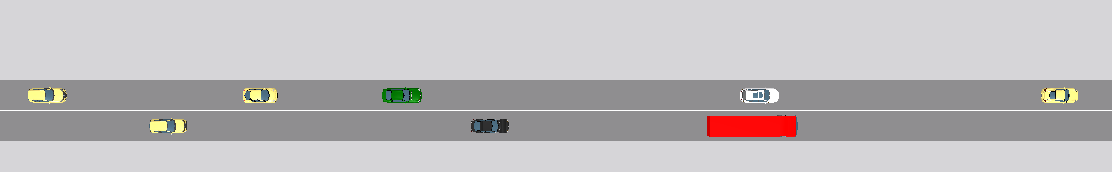}
    \caption{Vissim Simulation Snapshot. The red vehicle represents the uncontrolled vehicle $U$, the black vehicle represents CAV $C$, the white and green vehicles represent CAVs $i^*$ and $i^*+1$ respectively, and yellow vehicles represent surrounding traffic.}
    \label{fig5:Vissim}
    \vspace*{-\baselineskip}    
\end{figure}

Our simulation setting consists of a straight two-lane highway segment with an allowable speed range of $v_{\min}=16\, m/s$ and $v_{\max}=33\, m/s$. We define a traffic flow of 6000 vehicles/hour within a 5000m highway stretch. The incoming traffic (the yellow vehicles depicted in Fig. \ref{fig5:Vissim}) is spawned with a desired speed of $v_d = 29 \, m/s$. Similarly, the inter-vehicle safe distance \eqref{eq:safety_constraint} is given by $\delta = 1.5\,m$ and headway parameter $\varphi$ drawn from a normal distribution $\mathcal{N}(0.6,\,0.04)\,s$. In order to simulate congestion generation, we spawn an uncontrolled vehicle $U$ (the red truck as depicted in Fig. \ref{fig5:Vissim}) travelling on the right-lane (slow lane) with a constant speed $v_U=16\,m/s$ throughout the simulation. The corresponding CAV $C$ is defined as vehicle $U$'s immediately following vehicle (defined as the black colored vehicle in Fig. \ref{fig5:Vissim}). For the maneuver start distance we select $d_{\text{start}}$ from $\mathcal{N}(70,10)\,m$ for every CAV $C$ initiating a maneuver. The control limits specified for every CAV are given by $u_{\min}=-7 \,m/s^2$ and $u_{\max}=3.3 \, m/s^2$. It is assumed that all CAVs in the simulation share the same control bounds.

\subsection{CAV C Longitudinal Maneuver}
\begin{figure*}[hpt]
    \begin{subfigure}{.33\textwidth}
      \centering
      \includegraphics[width=.9\linewidth]{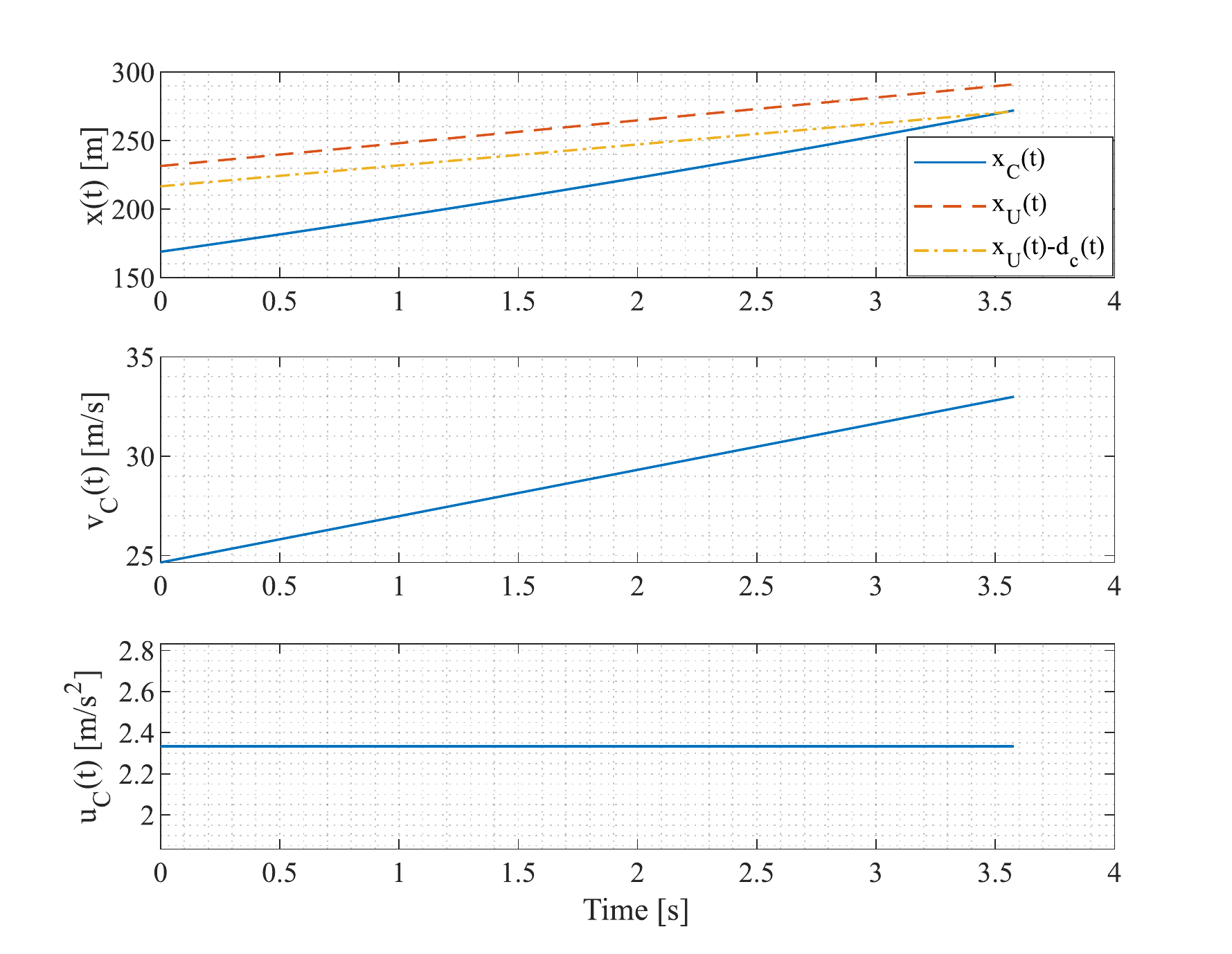}  
      \caption{\centering{Case 1: Constant Acceleration Sample for CAV $C$ with no time relaxation}}
      \label{fig6a:acceleration}
    \end{subfigure}
    \begin{subfigure}{.33\textwidth}
      \centering
      \includegraphics[width=.9\linewidth]{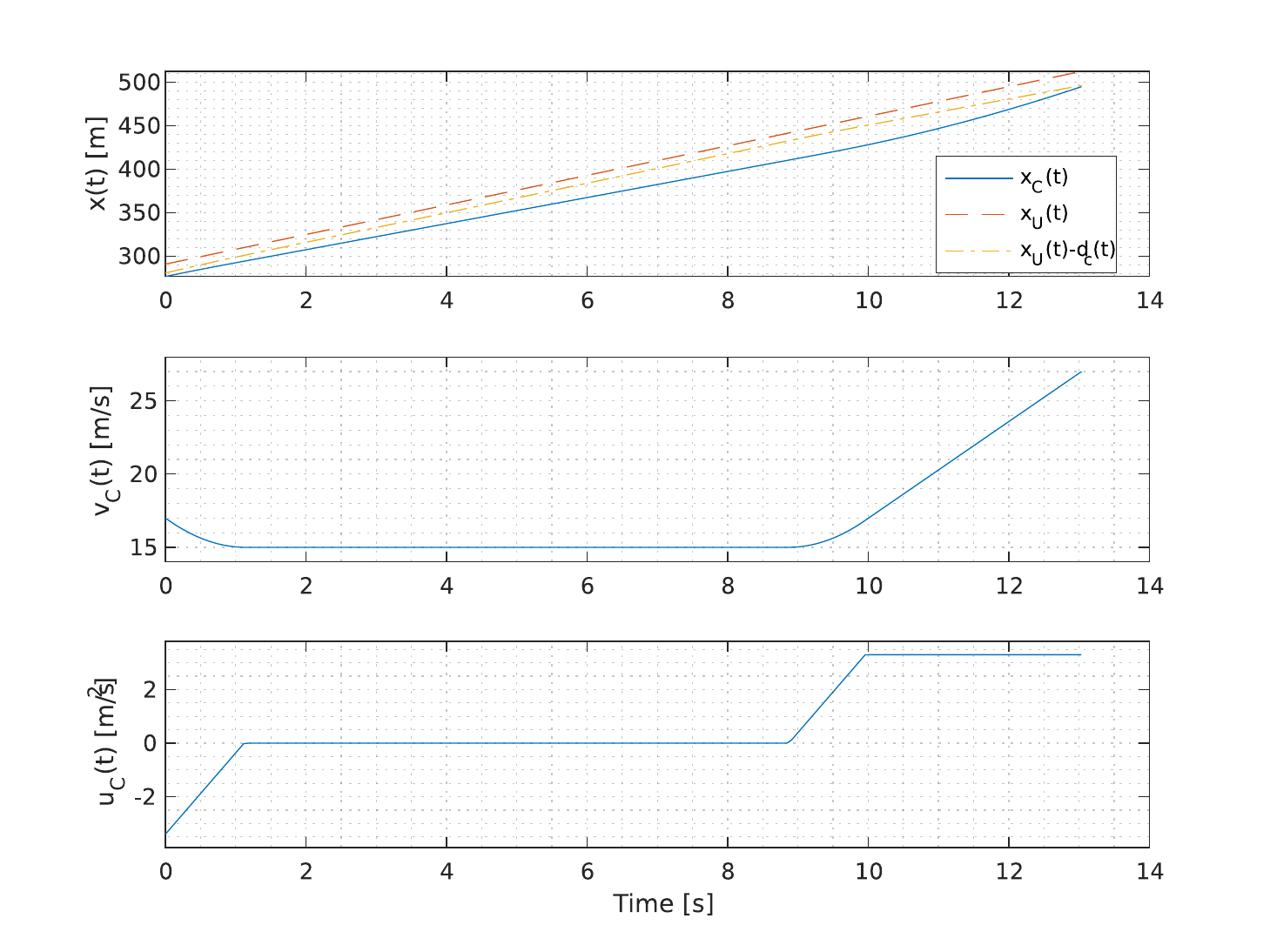}  
      \caption{\centering{Case 2: Mixed Acceleration Sample for CAV $C$ with no time relaxation}}
      \label{fig6b:close_acceleration}
    \end{subfigure}
    \begin{subfigure}{.33\textwidth}
      \centering
      \includegraphics[width=.9\linewidth]{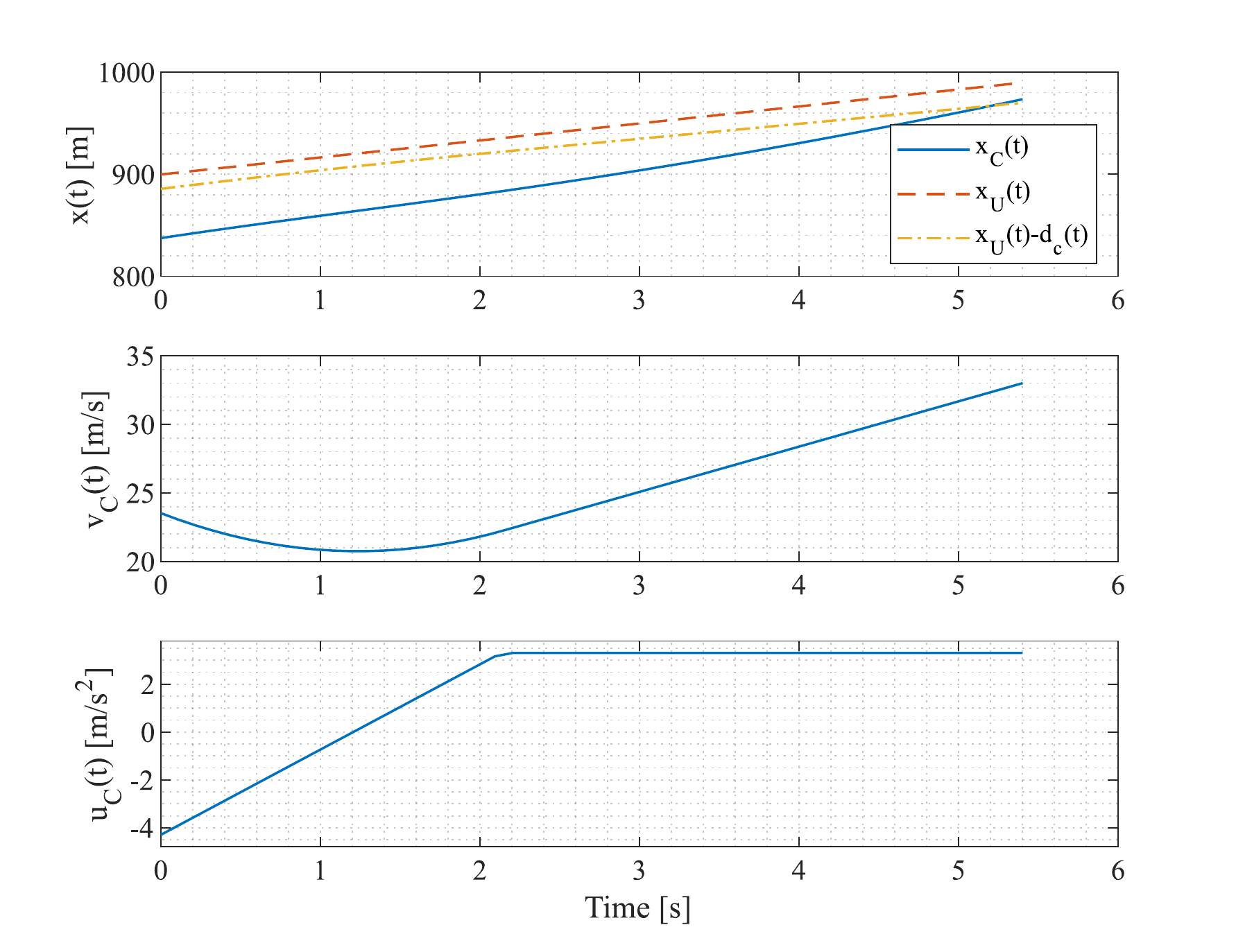}  
      \caption{\centering{Case 3:Mixed Acceleration Sample for CAV $C$ under time relaxation}}
      \label{fig6c:mixed_acceleration}
    \end{subfigure}
    \caption{Sample Optimal Trajectory Solutions for CAV $C$}
    \label{fig:CAV_C_AccelerationSamples}
\end{figure*}

We apply the formulation proposed for CAV $C$ in \eqref{eq:cav_c_objective_simplified} for different settings using a constant weight factor $\alpha=0.4$, desired speed $v_\text{d}=29\,m/s$, and a relaxation constant $\delta_\text{tol}=4\, m^2/s^4$. Thus, we provide simulation results for different initial conditions pertaining only to vehicles $U$ and $C$ as described in Table \ref{tab:vehicleCSample} for Cases 1 and 2. Similarly, for Case 3 we show a sample trajectory generated from the relaxation of the optimal time proposed in \eqref{eq:cav_c_relaxedTime}.  It can be seen for Cases 1 and 2 that the optimal terminal time solution $t_f^*$ inversely varies with the starting distance $d_\text{start}$. Thus, for Case 1, when  $d_\text{start}=70\,m$, the resulting maneuver time is $t^*_f=3.58\,s$ and a control strategy of constant acceleration to reach $v_\text{d}$ is shown in Fig. \ref{fig6a:acceleration}. Conversely, for Case 2, when $d_\text{start}=14\,m$, the resulting maneuver time is $t^*_f=13.03\,s$ given that CAV $C$ needs to first undergo a deceleration segment followed by a segment with no acceleration to provide enough space to undergo
a final acceleration
segment allowing CAV $C$ to reach $v_\text{d}$ without violating the safety constraint \eqref{eq4a:cavC_vehU_safety_constraint} as shown in Fig. \ref{fig6b:close_acceleration}. Lastly, for Case 3,  a relaxed terminal time constraint is provided with $t'_f=14.65\,s$. It can be seen from Fig. \ref{fig6c:mixed_acceleration} that the resulting optimal trajectory of CAV $C$, similar to Case 2, is composed of a negative acceleration segment followed by constant acceleration. 

\begin{table}[hpt]
\centering
\caption{Vehicle C Sample Results}
\label{tab:vehicleCSample}
\resizebox{\textwidth/2-6pt}{!}{%
\begin{tabular}{|c|c|c|c|c|c|c|c|c|}
\hline
    \diagbox[width=10em]{\textbf{{\ul Description}}}{\textbf{{\ul States}}} &
      \textbf{{\ul Relaxed}} &
      \textbf{\begin{tabular}[c]{@{}c@{}}$d_\text{start}$ \\ {[}\textit{m}{]}\end{tabular}} &
      \textbf{\begin{tabular}[c]{@{}c@{}}$x_U(t_0)$\\ {[}\textit{m}{]}\end{tabular}} &
      \textbf{\begin{tabular}[c]{@{}c@{}}$v_U(t_0)$\\ {[}\textit{m/s}{]}\end{tabular}} &
      \textbf{\begin{tabular}[c]{@{}c@{}}$x_C(t_0)$\\ {[}\textit{m}{]}\end{tabular}} &
      \textbf{\begin{tabular}[c]{@{}c@{}}$v_C(t_0)$\\ {[}\textit{m/s}{]}\end{tabular}} &
      \textbf{\begin{tabular}[c]{@{}c@{}}$t_f$\\ {[}\textit{s}{]}\end{tabular}} &
      \textbf{\begin{tabular}[c]{@{}c@{}}$v_C(t_f)$\\ {[}\textit{m/s}{]}\end{tabular}} \\ \hline
    Case 1 & False & 70 & 342 & 16 & 272 & 25 & 3.58  & 30.9 \\ \hline
    Case 2 & False & 14 & 290 & 16 & 272 & 17 & 13.03 & 27   \\ \hline
    Case 3 & True & 70 & 935 & 16 & 865 & 23 & 14.65 & 30.6 \\ \hline
    \end{tabular}%
    }
     \vspace*{-\baselineskip} 
\end{table}
\subsection{Sequential Maneuvers}
 We have also implemented a series of system-centric (social) optimal maneuvers. For this purpose, we discretize the start of the maneuvers as explained in Section \ref{SubSec3_F:Sequential_Maneuvers}. Additionally, a study was performed to determine the optimal parameters that would lead to minimal energy, time, and disruption, along with throughput improvements. Table \ref{tab:throughput_analysis} summarizes the results of our throughput study comparing the throughput under no cooperation with the case of vehicle-centric (selfish) maneuvers as in \cite{chen2020cooperative} and with the system-centric maneuvers presented in this paper with different parameters. The throughput analysis is performed by counting the number of vehicles within a $120\,s$ window that cross a measurement point located at $2000\, m$ from the starting line. The maximum time and maximum disruption values used for this study were given as $T_\text{th}=12\,s$ and $D_\text{th}=25\,m^2$  respectively. It is worth observing that placing much more emphasis on CAV $i^*+1$ as opposed to $i^*$ in the disruption metric (\ref{eq:disruption}) improves the throughput ($35\%$ over the no-cooperation case). This is consistent with the intuition that it is the CAV that decelerates to accommodate $C$ which may cause the most significant traffic flow disruption in the fast lane.

\begin{table}[hpt]
\centering
\caption{\centering{Throughput Simulation Study: Throughput results over 120 s analysis window}}
\label{tab:throughput_analysis}
\resizebox{\textwidth/2-6pt}{!}{%
\begin{tabular}{|c|c|c|c|c|c|c|}
\hline
{\ul \textbf{Description}} &
  {\ul \textbf{Relaxation}} &
  {\ul \textbf{$1-\gamma$}} &
  \textbf{\begin{tabular}[c]{@{}c@{}}Vehicle \\ Count\end{tabular}} &
  \textbf{\begin{tabular}[c]{@{}c@{}}Vehicle Flow \\ {[}\textit{veh/hour}{]}\end{tabular}} &
  \textbf{\begin{tabular}[c]{@{}c@{}}Avg. Travel \\ Time {[}\textit{s}{]}\end{tabular}} &
  \textbf{\begin{tabular}[c]{@{}c@{}}Avg Speed \\ {[}\textit{m/s}{]}\end{tabular}} \\ \hline
{\color[HTML]{FE0000} \textit{System-centric}} &
  {\color[HTML]{FE0000} T} &
  {\color[HTML]{FE0000} 0.99} &
  {\color[HTML]{FE0000} 50} &
  {\color[HTML]{FE0000} 1500} &
  {\color[HTML]{FE0000} 76.7} &
  {\color[HTML]{FE0000} 22.5} \\ \hline
\textit{System-centric} & T & 0.8 & 44 & 1320 & 78.1 & 21.8 \\ \hline
\textit{System-centric} & T & 0.5 & 44 & 1320 & 78.3 & 19.8 \\ \hline
\textit{System-centric} & F & 0.5 & 45 & 1350 & 77.8 & 19.8 \\ \hline
\textit{Vehicle-centric}      & F & 0.5 & 27 & 810  & 82.6 & 18.9 \\ \hline
\textit{No Cooperation}  & - & -   & 37 & 1110 & 85.4 & 19.7 \\ \hline
\end{tabular}%
}
 \vspace*{-\baselineskip}
\end{table}

%% file: sections/Conclusions.tex
We have developed a ``system-centric'' decentralized optimal control framework for multiple cooperating CAVs that minimizes energy and the maneuver time while also selecting an optimal cooperation pair of CAVs within a neighboring candidate set that minimizes a disruption metric for the fast lane traffic flow to ensure it never exceeds a given threshold. Our framework allows a relaxation of the minimum maneuver time to increase the chance that the disruption threshold constraint is satisfied, thus ensuring that throughput is improved. Our approach is extended to a discrete sequential maneuver process. Simulation results show the effectiveness of the proposed method with improvements of 35\% in throughput over the no-cooperation case. 

Ongoing work aims to perform  multiple maneuvers simultaneously while still minimizing the traffic disruption. Future work will include the incorporation of a ``comfort'' factor in the problem by minimizing any jerk resulting from accelerating/decelerating vehicles during the maneuvers. We are also working towards extending our analysis to a mixed traffic setting with both CAVs and human-driven vehicles and to include the stochastic behavior of uncontrolled vehicles while maintaining safety for all vehicles involved in a lane-changing maneuver.

%% file: root.bbl
\newcommand{\noopsort}[1]{} \newcommand{\printfirst}[2]{#1}
  \newcommand{\singleletter}[1]{#1} \newcommand{\switchargs}[2]{#2#1}
\begin{thebibliography}{10}
\providecommand{\url}[1]{#1}
\csname url@samestyle\endcsname
\providecommand{\newblock}{\relax}
\providecommand{\bibinfo}[2]{#2}
\providecommand{\BIBentrySTDinterwordspacing}{\spaceskip=0pt\relax}
\providecommand{\BIBentryALTinterwordstretchfactor}{4}
\providecommand{\BIBentryALTinterwordspacing}{\spaceskip=\fontdimen2\font plus
\BIBentryALTinterwordstretchfactor\fontdimen3\font minus
  \fontdimen4\font\relax}
\providecommand{\BIBforeignlanguage}[2]{{%
\expandafter\ifx\csname l@#1\endcsname\relax
\typeout{** WARNING: IEEEtran.bst: No hyphenation pattern has been}%
\typeout{** loaded for the language `#1'. Using the pattern for}%
\typeout{** the default language instead.}%
\else
\language=\csname l@#1\endcsname
\fi
#2}}
\providecommand{\BIBdecl}{\relax}
\BIBdecl

\bibitem{varaiya1993smart}
P.~Varaiya, ``Smart cars on smart roads: problems of control,'' \emph{IEEE
  Trans. on Automatic Control}, vol.~38, no.~2, pp. 195--207, 1993.

\bibitem{zhao2018accelerated}
D.~Zhao, X.~Huang, H.~Peng, H.~Lam, and D.~J. LeBlanc, ``Accelerated evaluation
  of automated vehicles in car-following maneuvers,'' \emph{IEEE Trans. on
  Intelligent Transportation Systems}, vol.~19, no.~3, pp. 733--744, 2018.

\bibitem{wang2016cooperative}
M.~Wang, W.~Daamen, S.~P. Hoogendoorn, and B.~van Arem, ``Cooperative
  car-following control: Distributed algorithm and impact on moving jam
  features,'' \emph{IEEE Trans. on Intelligent Transportation Systems},
  vol.~17, no.~5, pp. 1459--1471, 2016.

\bibitem{wang2015game}
M.~Wang, S.~P. Hoogendoorn, W.~Daamen, B.~van Arem, and R.~Happee, ``Game
  theoretic approach for predictive lane-changing and car-following control,''
  \emph{Transportation Research Part C: Emerging Technologies}, vol.~58, pp.
  73--92, 2015.

\bibitem{fleck2015adaptive}
J.~L. Fleck, C.~G. Cassandras, and Y.~Geng, ``Adaptive quasi-dynamic traffic
  light control,'' \emph{IEEE Transactions on Control Systems Technology},
  vol.~24, no.~3, pp. 830--842, 2015.

\bibitem{dresner2008multiagent}
K.~Dresner and P.~Stone, ``A multiagent approach to autonomous intersection
  management,'' \emph{{Journal of Artificial Intelligence Research}}, vol.~31,
  pp. 591--656, 2008.

\bibitem{zhang2019decentralized}
Y.~Zhang and C.~G. Cassandras, ``Decentralized optimal control of connected
  automated vehicles at signal-free intersections including comfort-constrained
  turns and safety guarantees,'' \emph{{Automatica}}, vol. 109, p. 108563,
  2019.

\bibitem{nilsson2015longitudinal}
J.~Nilsson, M.~Br{\"a}nnstr{\"o}m, E.~Coelingh, and J.~Fredriksson,
  ``Longitudinal and lateral control for automated lane change maneuvers,''
  \emph{Proc. of 2015 American Control Conf.}, pp. 1399--1404, 2015.

\bibitem{bax2014road}
C.~Bax, P.~Leroy, and M.~P. Hagenzieker, ``Road safety knowledge and policy: A
  historical institutional analysis of the {Netherlands},''
  \emph{Transportation Research part F: Traffic Psychology and Behaviour},
  vol.~25, pp. 127--136, 2014.

\bibitem{you2015trajectory}
F.~You, R.~Zhang, G.~Lie, H.~Wang, H.~Wen, and J.~Xu, ``Trajectory planning and
  tracking control for autonomous lane change maneuver based on the cooperative
  vehicle infrastructure system,'' \emph{Expert Systems with Applications},
  vol.~42, no.~14, pp. 5932--5946, 2015.

\bibitem{werling2010optimal}
M.~Werling, J.~Ziegler, S.~Kammel, and S.~Thrun, ``Optimal trajectory
  generation for dynamic street scenarios in a frenet frame,'' \emph{Proc. of
  2010 IEEE Intl. Conf. on Robotics and Automation}, pp. 987--993, 2010.

\bibitem{bevly2016lane}
D.~Bevly, X.~Cao, M.~Gordon, G.~Ozbilgin, D.~Kari, B.~Nelson, J.~Woodruff,
  M.~Barth, C.~Murray, A.~Kurt \emph{et~al.}, ``Lane change and merge maneuvers
  for connected and automated vehicles: A survey,'' \emph{IEEE Trans. on
  Intelligent Vehicles}, vol.~1, no.~1, pp. 105--120, 2016.

\bibitem{nilsson2017lane}
J.~Nilsson, M.~Br{\"a}nnstr{\"o}m, E.~Coelingh, and J.~Fredriksson, ``Lane
  change maneuvers for automated vehicles,'' \emph{IEEE Trans. on Intelligent
  Transportation Systems}, vol.~18, no.~5, pp. 1087--1096, 2017.

\bibitem{mahjoub2017learning}
H.~N. Mahjoub, A.~Tahmasbi-Sarvestani, H.~Kazemi, and Y.~P. Fallah, ``A
  learning-based framework for two-dimensional vehicle maneuver prediction over
  v2v networks,'' \emph{Proc. of 15th IEEE Intl. Conf. on Dependable, Autonomic
  and Secure Computing}, pp. 156--163, 2017.

\bibitem{desiraju2014minimizing}
D.~Desiraju, T.~Chantem, and K.~Heaslip, ``Minimizing the disruption of traffic
  flow of automated vehicles during lane changes,'' \emph{{Proc.\ IEEE Int.\
  Conf.\ on Intelligent Transportation Systems}}, vol.~16, no.~3, pp.
  1249--1258, 2015.

\bibitem{luo2016dynamic}
``A dynamic automated lane change maneuver based on vehicle-to-vehicle
  communication,'' \emph{Transportation Research Part C: Emerging
  Technologies}, vol.~62, pp. 87--102, 2016.

\bibitem{li2020cooperative}
T.~Li, J.~Wu, C.-Y. Chan, M.~Liu, C.~Zhu, W.~Lu, and K.~Hu, ``A cooperative
  lane change model for connected and automated vehicles,'' \emph{{IEEE
  Access}}, vol.~8.

\bibitem{kamal2013model}
M.~A.~S. Kamal, M.~Mukai, J.~Murata, and T.~Kawabe, ``Model predictive control
  of vehicles on urban roads for improved fuel economy,'' \emph{IEEE Trans. on
  Control Systems Technology}, vol.~21, no.~3, pp. 831--841, 2013.

\bibitem{katriniok2013optimal}
A.~Katriniok, J.~P. Maschuw, F.~Christen, L.~Eckstein, and D.~Abel, ``Optimal
  vehicle dynamics control for combined longitudinal and lateral autonomous
  vehicle guidance,'' \emph{Proc. of 2013 Control Conf.}, pp. 974--979, 2013.

\bibitem{li2018balancing}
B.~Li, Y.~Zhang, Y.~Feng, Y.~Zhang, Y.~Ge, and Z.~Shao, ``Balancing computation
  speed and quality: A decentralized motion planning method for cooperative
  lane changes of connected and automated vehicles,'' \emph{{IEEE Transactions
  on Intelligent Vehicles}}, vol.~3, no.~3, pp. 340--350, 2018.

\bibitem{lam2013cooperative}
S.~Lam and J.~Katupitiya, ``Cooperative autonomous platoon maneuvers on
  highways,'' \emph{Proc. of 2013 IEEE/ASME Intl. Conf. on Advanced Intelligent
  Mechatronics}, pp. 1152--1157, 2013.

\bibitem{li2017optimal}
B.~Li, Y.~Zhang, Y.~Ge, Z.~Shao, and P.~Li, ``Optimal control-based online
  motion planning for cooperative lane changes of connected and automated
  vehicles,'' \emph{Proc. of 2017 IEEE/RSJ Intl. Conf. on Intelligent Robots
  and Systems}, pp. 3689--3694, 2017.

\bibitem{chen2020cooperative}
R.~Chen, C.~G. Cassandras, A.~Tahmasbi-Sarvestani, S.~Saigusa, H.~N. Mahjoub,
  and Y.~K. Al-Nadawi, ``Cooperative time and energy-optimal lane change
  maneuvers for connected automated vehicles,'' \emph{{IEEE Transactions on
  Intelligent Transportation Systems}}, 2020.

\bibitem{vogel2003comparison}
K.~Vogel, ``A comparison of headway and time to collision as safety
  indicators,'' \emph{Accident Analysis \& Prevention}, vol.~35, no.~3, pp.
  427--433, 2003.

\end{thebibliography}
